\documentclass{article}

\usepackage{PRIMEarxiv}
\usepackage{amsmath}
\usepackage{amssymb}
\usepackage[utf8]{inputenc} 
\usepackage{helvet}         
\usepackage[hidelinks]{hyperref} 
\usepackage{url}            
\usepackage{booktabs}       
\usepackage{amsfonts}       
\usepackage{nicefrac}       
\usepackage{microtype}      
\usepackage{lipsum}
\usepackage{fancyhdr}       
\usepackage{graphicx}       
\graphicspath{{media/}}     

\usepackage{braket}
\usepackage{multirow}
\usepackage{xcolor}
\usepackage{listings}
\usepackage{quantikz}
\usepackage{tikz}
\usepackage{soul}
\usepackage{algorithm}
\usepackage{algpseudocode}
\usepackage{comment}
\usepackage{enumitem}

\usetikzlibrary{arrows.meta,positioning,calc,fit}

\sethlcolor{yellow}

\definecolor{codegreen}{rgb}{0,0.6,0}
\definecolor{codegray}{rgb}{0.5,0.5,0.5}
\definecolor{codepurple}{rgb}{0.58,0,0.82}
\definecolor{backcolour}{rgb}{0.95,0.95,0.92}

\lstdefinestyle{Pythonstyle}{
    language=Python,
    backgroundcolor=\color{backcolour},
    commentstyle=\color{codegreen},
    keywordstyle=\color{magenta},
    numberstyle=\tiny\color{codegray},
    stringstyle=\color{codepurple},
    basicstyle=\ttfamily\footnotesize,
    breakatwhitespace=false,
    breaklines=true,
    captionpos=b,
    keepspaces=true,
    numbers=left,
    numbersep=5pt,
    showspaces=false,
    showtabs=false,
    tabsize=2
}

\title{The Evolution of IBM's Quantum Information Software Kit (\textit{Qiskit}): A Review of its Applications}

\author{
  Param Pathak$^{1}$ \\
  QuantumAI Lab, Fractal Analytics \\
  Mumbai, Maharashtra, India \\
  \texttt{param.pathak@fractal.ai} \\
  \AND
  K. Tarakeshwar$^{2}$ \\
  Department of Computer Science \\
  Birla Institute of Technology and Science Pilani, Dubai Campus \\
  Dubai International Academic City, Dubai -- 345055, UAE \\
  \texttt{f20230241@dubai.bits-pilani.ac.in} \\
  \AND
  Syed Sufiyan Ali$^{2}$ \\
  Department of Computer Science \\
  Birla Institute of Technology and Science Pilani, Dubai Campus \\
  Dubai International Academic City, Dubai -- 345055, UAE \\
  \texttt{f20230258@dubai.bits-pilani.ac.in} \\
  \AND
  Shalini Devendrababu$^{1}$ \\
  QuantumAI Lab, Fractal Analytics \\
  Gurugram, Haryana, India \\
  \texttt{shalini.devendrababu@fractal.ai} \\
  \AND
  Adarsh Ganesan$^{3,4}$ \thanks{Corresponding author. ORCID: 0000-0002-5107-8452} \\
  Department of Electrical and Electronics Engineering \\
  Birla Institute of Technology and Science Pilani, Dubai Campus \\
  Dubai International Academic City, Dubai -- 345055, UAE \\
  Department of Mechanical Engineering \\
  Birla Institute of Technology and Science Pilani, Pilani Campus \\
  Pilani, Rajasthan, India \\
  \texttt{adarsh@dubai.bits-pilani.ac.in} \\
}

\begin{document}
\maketitle

\begin{abstract}
Quantum computing is being increasingly explored for computationally demanding
problems across various domains. However, the availability of accessible and
scalable software frameworks remains essential for practical experimentation and
adoption. IBM's open source quantum computing toolkit \textit{Qiskit} has become
an important framework in this space by offering tools for circuit design,
simulation, hardware execution, and domain specific applications. This review
examines how \textit{Qiskit} has evolved and what it has contributed to several
critical fields including cryptography and cybersecurity, image and signal
processing, climate science and energy applications, and finance and risk
analysis. We show how \textit{Qiskit} facilitates hybrid classical quantum
workflows and enables the execution of algorithms on physical quantum hardware,
while distinguishing results from ideal simulation, noisy simulation, and
hardware experiments. The review also traces the evolution of \textit{Qiskit}'s
architecture and APIs, and the consequences of these changes for the
compatibility of previously published workflows with contemporary versions of the
framework. Our exploration covers applications including quantum key
distribution, environmental modelling, energy optimization, portfolio
optimization, and financial risk analysis. By bringing together developments
scattered across different areas, this work identifies current limitations in
scalability, hardware noise, and software compatibility, and provides a reference
for researchers and practitioners working with \textit{Qiskit}.
\end{abstract}


\section{Introduction}\label{sec:intro}

Quantum computing (QC)~\cite{nielsen2010quantum} offers transformative
potential for fields including cryptography and materials science by
addressing problems that classical computers cannot feasibly
solve~\cite{shor1999polynomial}. Algorithms such as Shor's factorization
method~\cite{shor1994algorithms} and Grover's search
algorithm~\cite{grover1996fast} promise substantial speed-ups for important
computational tasks. Realising practical quantum advantage, however, is not
solely a hardware problem. It also requires software frameworks that are
usable in practice and that connect algorithmic theory to the devices on
which algorithms must actually run.

\textit{Qiskit}~\cite{javadi2024quantum}, International Business Machines' (IBM) open-source quantum
computing software development kit (SDK), has become a central instrument in that
connection. It abstracts hardware-specific detail and provides modular tools
for circuit construction, compilation, simulation and execution, giving
researchers, developers and industrial users a common route to quantum
resources~\cite{fingerhuth2018open}. Domain-specific packages extend this
core into chemistry, optimization, machine learning (ML) and
finance~\cite{sahin2025qiskit}, and the framework now supports work in
research~\cite{stirbu2023quantum}, education~\cite{seskir2022quantum} and
industrial application~\cite{alexander2020qiskit}. Its coupling to IBM's
hardware programme, which currently spans the Heron processor family and the
120-qubit Nighthawk architecture~\cite{ibmhardware} and is directed toward the
fault-tolerant Starling system~\cite{ibmroadmap}, positions it between
algorithmic theory and the devices available to researchers
today~\cite{shammah2023open}.

That centrality is precisely what makes the framework difficult to survey.
\textit{Qiskit} has been restructured repeatedly since 2017 in its packaging, execution model, and internal implementation, while much of the published
application literature was developed using interfaces that have since been changed or removed. As a result, some workflows reported in earlier studies
cannot be executed under contemporary \textit{Qiskit} versions without modification. Accounting for this evolution is therefore important when interpreting the
application literature, particularly when comparing implementations developed at different stages of the framework's development.

This review is organized around four questions:

\begin{itemize}
\item[\textbf{RQ1}] How has \textit{Qiskit}'s architecture and programming model evolved from its original element structure to the contemporary 2.x ecosystem?
\item[\textbf{RQ2}] Which application problems and quantum algorithm families are represented in the \textit{Qiskit} literature across the domains surveyed here?
\item[\textbf{RQ3}] What level of experimental evidence supports these applications in terms of execution mode, problem scale, treatment of noise, error mitigation, and comparison with classical methods?
\item[\textbf{RQ4}] Which software, hardware, and methodological limitations currently constrain \textit{Qiskit} based demonstrations from progressing toward practically meaningful problem sizes?
\end{itemize}

\noindent
In addressing these questions, this work makes four contributions:

\begin{itemize}

\item A version-anchored account of \textit{Qiskit}'s development from 2017
to 2026, distinguishing the historical element architecture from the
contemporary stack, and identifying the points at which interface changes broke
compatibility with existing code.

\item An examination of the surveyed application literature that applies a
consistent set of questions across all four domains, considering the problem
addressed, the quantum method used, the role of \textit{Qiskit} in the
implementation, and the level of experimental evidence reported. Results
obtained from ideal simulation, noisy simulation, and physical hardware are
distinguished where reported, together with the problem scale, treatment of noise,
error mitigation where applied, the classical comparisons reported by each
study, and the \textit{Qiskit} or ecosystem package versions where these are
stated.

\item A set of worked implementations that express the circuit constructions
discussed in each domain using the interfaces available in the reference version
of \textit{Qiskit}, in place of the earlier interfaces used by several of the
original studies.

\item A critical assessment of where \textit{Qiskit}-based work has produced
convincing evidence and where it remains at the level of demonstration.
\end{itemize}

The applications surveyed span four interdisciplinary areas: cryptography
and cybersecurity, image and signal processing, climate science and energy
applications, and finance and risk analysis. Across these domains, the
review examines the problems addressed, the quantum methods used, the role
of \textit{Qiskit} in implementation, and the experimental evidence reported in the
literature.

The remainder of the paper is organized as follows.
Section~\ref{sec:evolution} traces the evolution of \textit{Qiskit} from 2017 to
2026 and establishes the version boundaries used throughout the review.
Section~\ref{sec:architecture} describes the contemporary \textit{Qiskit}
architecture and execution workflow. Section~\ref{sec:applications} examines
the use of \textit{Qiskit} across the four application domains and distinguishes
evidence obtained from ideal simulation, noisy simulation, and physical
quantum hardware. Within these domains, the reviewed studies are considered
in terms of their problem scale, implementation approach, treatment of
noise, classical comparisons, and reported limitations. The application
analysis covers quantum key distribution (QKD) and random number generation,
medical imaging and telecommunications, climate modelling and energy
optimization, and portfolio optimization and financial risk analysis.
Section~\ref{sec:conclusion} brings together the main findings from the
evolution, architecture, and application literature and concludes the
paper.

\section{The Evolution of \textit{Qiskit}, 2017--2026} \label{sec:evolution}

A review of a software framework is only as current as the version it
describes. \textit{Qiskit} has been under continuous development since March 2017,
and over that period its packaging, its execution model and even its
internal implementation language have been restructured more than once.
Descriptions of the framework written at different points in its history are
therefore not merely more or less detailed; in several respects they are
mutually incompatible. Many common programming patterns used in earlier
\textit{Qiskit} releases are incompatible with the contemporary interface, and
architectural accounts that were accurate when published are now misleading.

This section traces \textit{Qiskit}'s development through five phases: the original element architecture (\ref{sub:elements}); the decomposition of that architecture (\ref{sub:decomposition}); the introduction of primitives as the execution model (\ref{sub:primitives}); the compatibility changes introduced with \textit{Qiskit}~1.0 (\ref{sub:onepoint}); and the development of the 2.x series around a Rust-based core and C application programming interface (API) (\ref{sub:twoseries}). These changes provide the historical context needed to interpret the application literature discussed in Section~\ref{sec:applications}.

\subsection{The element architecture (2017--2021)} \label{sub:elements}

\textit{Qiskit} was first released in March 2017 and was for several years organized
as a \emph{metapackage}: the distribution installed by \texttt{pip install
qiskit} was a thin wrapper that pulled in a set of separately versioned
components, referred to as \emph{elements}. Four elements defined the
framework during this period.

\textit{Terra} provided the core data model: quantum circuits, operators,
registers, together with the transpiler, scheduling, and the abstract
backend interface. \textit{Aer} supplied high-performance simulators,
including statevector, density-matrix, stabilizer and matrix-product-state
methods, along with configurable noise models. \textit{Ignis} offered tools
for noise characterization, calibration and error mitigation, covering
randomized benchmarking, quantum volume, and state and process tomography.
\textit{Aqua} sat above these and supplied application-level algorithms,
organized into chemistry, optimization, ML and finance
sub-domains.

Two properties of this arrangement matter for what followed. First, the
elements were versioned independently but distributed together, so a single
metapackage version number corresponded to a specific combination of element
versions --- a mapping that the release notes tracked in a dedicated
table~\cite{qiskit044notes}. Second, execution was expressed through a
free function, \texttt{execute(circuit, backend, shots=...)}, which
compiled and submitted a circuit in one step and returned a result object
containing measurement counts. This function, and the \texttt{Aer.get\_backend()}
call typically used to obtain the backend, are the signature of code written
in this era.

\subsection{Decomposition of the elements (2021--2023)} \label{sub:decomposition}

The element architecture was dismantled between 2021 and 2023, element by
element.

\textbf{Aqua} was deprecated first. Version 0.9.0, released on 2 April 2021,
marked the package as deprecated and began the migration of its
contents~\cite{qiskitaqua}. The core algorithm and operator machinery moved
into Terra as \texttt{qiskit.algorithms} and \texttt{qiskit.opflow}, while
the application layers were separated into standalone repositories:
\textit{Qiskit} Nature, \textit{Qiskit} Optimization, \textit{Qiskit} ML and \textit{Qiskit}
Finance. This is the change that produced the domain-specific packages that
most application papers now depend on, and it is worth emphasizing that
those packages exist \emph{because} Aqua was dissolved, not alongside it.

\textbf{Ignis} was deprecated with \textit{Qiskit} 0.33 (Terra 0.19.1) and
superseded by \textit{Qiskit} Experiments, with active development ending at
Ignis 0.7.0 ~\cite{qiskit033notes,qiskitignis}. Its functionality was
redistributed rather than migrated as a single package. Characterization
and calibration tools moved primarily to
\path{qiskit_experiments.library.characterization} and
\path{qiskit_experiments.library.calibration}, while randomized
benchmarking, quantum volume, and tomography became part of \textit{Qiskit}
Experiments. Readout-mitigation functionality moved elsewhere in the
\textit{Qiskit} ecosystem, and some components, including \texttt{ZZFitter}, were
not migrated. Consequently, Ignis should be treated as part of \textit{Qiskit}'s
historical architecture rather than its contemporary software stack.

\textbf{The provider} followed. \textit{Qiskit} 0.40 (January 2023) formally
deprecated \texttt{qiskit-ibmq-provider} within the
metapackage ~\cite{qiskit040notes}, part of a stated move toward a model in
which the \texttt{qiskit} package contains only core circuit-construction and compilation functionality, with hardware and simulator links installable separately. Cloud access migrated to \texttt{qiskit-ibm-runtime} and
\texttt{qiskit-ibm-provider}.

\textbf{The metapackage itself} was retired last. \textit{Qiskit} 0.44 (July 2023)
was the final metapackage release, and removed \texttt{qiskit-aer} and
\texttt{qiskit-ibmq-provider} from the distribution ~\cite{qiskit044notes};
\textit{Qiskit} 0.45 was the first release after the element structure had been
removed entirely ~\cite{qiskit045notes}. From this point what had been called
``\textit{Qiskit} Terra'' is simply called \textit{Qiskit}.

\subsection{Primitives as the execution model (2022 to 2024)} \label{sub:primitives}

Alongside the changes in \textit{Qiskit}'s packaging structure, the way quantum
programs were executed also changed considerably.

In the element era, users typically submitted a circuit to a backend and
received measurement counts as the result. The primitives model introduced
a different approach through two main abstractions. The \texttt{Sampler}
returns data derived from measurement outcomes, while the \texttt{Estimator}
returns expectation values of specified observables. This is particularly
useful for variational algorithms, which form a large part of the application
literature and are naturally expressed in terms of expectation values. Earlier
workflows often required users to reconstruct these quantities manually from
measurement counts. Primitives instead place this functionality behind a
common execution interface, together with support for error mitigation in relevant implementations.

The primitives were later revised. The V2 interfaces became available with
\texttt{qiskit-ibm-runtime} 0.21.0 and introduced the Primitive Unified Bloc
(PUB) input format together with array valued results. The revision also
changed what the \texttt{Sampler} returns: the V1 interface produced quasi
probability distributions, whereas the V2 interface returns the individual
measurement outcomes of each shot. The V1 \texttt{Sampler} and
\texttt{Estimator} were deprecated in \texttt{qiskit-ibm-runtime} 0.23 on 15
April 2024, and support was removed on 15 August 2024
~\cite{qiskitv2primitives}. As a result, workflows written for the V1
primitives, including some developed after \textit{Qiskit} 1.0, also need to be
migrated.

The transition from V1 to V2 also shows that the primitives interface has
continued to evolve. The execution model has changed further during the
\textit{Qiskit} 2.x period, so the description given here reflects the framework at
the version considered in this review rather than a permanently fixed
design.

\subsection{Qiskit 1.0: the compatibility break} \label{sub:onepoint}

\textit{Qiskit} 1.0 was released on 15 February 2024, following a prerelease on
4 December 2023 and the transitional \textit{Qiskit} 0.46 release, which included
deprecation warnings for functionality scheduled to change in \textit{Qiskit} 1.0 ~\cite{qiskit1launch,qiskit046notes}. The release introduced several changes that are important when interpreting code written for
earlier versions of \textit{Qiskit}.

The metapackage structure was removed, and the
\texttt{qiskit.algorithms} module was separated from the core \textit{Qiskit}
package. The \texttt{qiskit.opflow} module, which had been widely used in
variational algorithm implementations, was also removed. At the same time,
primitives became the main interface for executing quantum workloads, and
existing functionality was reorganized around this model
\cite{qiskit1launch}. Because \textit{Qiskit} 1.0 used a packaging structure that
was incompatible with previous 0.x installations, users were advised to
create a new virtual environment rather than upgrade an existing
installation directly.

\textit{Qiskit} 1.0 also introduced several improvements to the underlying software.
The continuing migration of internal components to Rust and changes to the
transpiler improved both execution speed and memory efficiency. IBM reported
that circuit binding and transpilation were up to 16$\times$ faster than in
\textit{Qiskit} 0.33, while average memory usage was reduced by 55\% compared with
\textit{Qiskit} 0.39 ~\cite{qiskit1launch}. The release also introduced a versioning
policy intended to reduce the frequency of breaking changes in later
versions.

The effect of these changes depends on the part of \textit{Qiskit} used by an
application. Basic circuit construction with \texttt{QuantumCircuit}
remained largely compatible, while several execution and algorithm
interfaces were removed. These include \texttt{execute()},
\texttt{QuantumInstance}, \texttt{qiskit.opflow}, and algorithm classes
that accepted a \texttt{quantum\_instance} argument. Code that depends on
these interfaces therefore requires modification before it can run under
\textit{Qiskit} 1.0 or later. Simulator access also changed. The
\texttt{qiskit.Aer} entry point was removed from the main \textit{Qiskit} namespace,
while Aer continued as a separately installed package. As a result,
\texttt{Aer.get\_backend()} remains available when Aer is imported from
\texttt{qiskit\_aer}. Older \textit{Qiskit} programs may therefore require changes
to imports, execution interfaces, or algorithm implementations depending on
the APIs they use ~\cite{qiskit1features}.

\subsection{The 2.x series: Rust core, C API and compilation for fault tolerance} \label{sub:twoseries}

The \textit{Qiskit} 2.x series introduced further changes to the compiler,
internal data model, and interfaces available to developers. These releases
also expanded support for using \textit{Qiskit} from environments other than Python
and introduced compilation tools designed for future fault tolerant
quantum systems.

\textit{Qiskit} 2.0 was released in March 2025, followed by \textit{Qiskit} 2.1 in June 2025
and \textit{Qiskit} 2.2 in September 2025. These releases maintained compatibility
within the 2.x series ~\cite{qiskitroadmap}. \textit{Qiskit} 2.3, released in January
2026, completed the migration of the internal data model to Rust, with
\texttt{ControlFlowOp} being the final major component transferred to the
new implementation. The release also expanded the C API and allowed custom
transpiler passes to be written in C using \texttt{QkDag} and
\texttt{QkTarget} ~\cite{qiskit23blog}.

\textit{Qiskit} 2.4 was released in April 2026 and added support for compiled Python
extensions based on the C API. It also introduced improvements to the
transpilation pipeline for fault tolerant circuits, including methods aimed
at reducing the number of $T$ gates required by compiled
circuits ~\cite{qiskit24blog}. \textit{Qiskit} 2.5 followed in July 2026 and added
preset pass managers for Pauli based computation and Clifford$+T$
representations. The Clifford$+T$ workflow is available through
\texttt{generate\_clifford\_t\_pass\_manager()} and provides a dedicated
compilation path for fault tolerant targets. The release also extended the
C API to support inspection of classical control flow and increased the
minimum supported versions to NumPy 2.0 and SciPy 1.14 under a dependency
policy based on the scientific Python community's SPEC~0
guidelines ~\cite{qiskit25blog}. Support for the \textit{Qiskit} 1.x series ended
with \textit{Qiskit} SDK 1.4.6 on 12 June 2026~\cite{qiskit25blog}.

Python continues to provide the main user interface to \textit{Qiskit}, while a growing part of the internal data model and other performance sensitive
components are implemented in Rust. The C API also allows \textit{Qiskit} functionality to be integrated into compiled applications and high performance computing environments where Python may not be the primary host language. At the same time, recent compiler development has included support for Clifford$+T$ synthesis, reduction of $T$ gate counts, and Pauli based computation. These developments are consistent with IBM's planned progression from current processors such as \textit{Nighthawk} and \textit{Loon} toward the fault tolerant Starling system ~\cite{ibmroadmap}.

Around the release of \textit{Qiskit} 2.5, IBM also renamed the \textit{Qiskit} Runtime
Service as the IBM Quantum Compute Service. The change was intended to
distinguish the open source \textit{Qiskit} SDK from IBM's managed cloud execution
service and did not require changes to existing APIs or workflows ~\cite{qiskit25blog}. Older publications may therefore use \textit{Qiskit} Runtime terminology when referring to the same managed execution service.

\section{Contemporary \textit{Qiskit} Architecture and Execution Workflow} \label{sec:architecture}

This section describes the current structure of \textit{Qiskit} using \textit{Qiskit} SDK 2.5
as the reference version. The specific version is stated because, as
discussed in Section~\ref{sec:evolution}, several components
and interfaces have changed considerably across \textit{Qiskit} releases.

The current \textit{Qiskit} environment consists of a relatively compact open source
core together with separately versioned packages and service interfaces.
For clarity, it can be divided into four main parts: the SDK, simulation,
execution, and the ecosystem of domain and utility packages.

\subsection{The SDK core: circuits, targets and compilation} \label{sub:sdk-core}

The \texttt{qiskit} package contains the main functionality required to
construct quantum circuits and operators and to compile them for a chosen
execution target. Other capabilities are provided through separately
installed packages.

\textbf{Circuit and operator construction} is centred on
\texttt{QuantumCircuit}, together with the circuit library and the
\texttt{quantum\_info} module for working with operators, states, and
channels. This part of \textit{Qiskit} remained relatively stable during the
transition to \textit{Qiskit} 1.0 compared with the execution and algorithm
interfaces.

\textbf{The \texttt{Target}} provides a description of the capabilities of
an execution backend. It replaces the earlier configuration and properties
objects with a single representation that specifies the supported
instructions, the qubits or qubit pairs on which those instructions can be
applied, and information such as gate error rates and operation durations.
The transpiler uses this information when adapting a circuit to the
constraints of a particular device. The same representation can also be
accessed through \textit{Qiskit}'s C interface.

\textbf{Compilation} is performed through a pass manager. The
\texttt{generate\_preset\_pass\_manager()} function creates a standard
compilation pipeline for a selected optimization level. The pipeline
includes layout selection, routing according to the device connectivity,
translation into the supported basis gates, and circuit optimization.
\textit{Qiskit} 2.5 also provides compilation pipelines intended for fault tolerant
targets. These include
\texttt{generate\_clifford\_t\_pass\_manager()} for Clifford$+T$
compilation and a separate pipeline for Pauli based computation \cite{qiskit25blog}.

\textbf{The C API} makes parts of the \textit{Qiskit} compiler available outside
Python. From \textit{Qiskit} 2.3, developers can implement custom transpiler passes
in C using \texttt{QkDag} and \texttt{QkTarget}. \textit{Qiskit} 2.4 extended this
support by allowing such components to be built and distributed as compiled
Python extensions ~\cite{qiskit23blog,qiskit24blog}. These capabilities make
it possible to integrate the \textit{Qiskit} compiler into larger high performance
computing workflows in which Python may not be the main programming
environment.

\subsection{Simulation} \label{sub:simulation}

Aer is distributed separately from the main \textit{Qiskit} SDK and is installed as
\texttt{qiskit-aer}, with the Python module imported as
\texttt{qiskit\_aer}. It provides several simulation methods used throughout
the application literature, including statevector, density matrix,
stabilizer, and matrix product state simulation. Aer also supports
configurable noise models, including models constructed using calibration
information obtained from physical quantum devices.

In this review, we distinguish between ideal simulation, simulation with a
noise model, and execution on physical quantum hardware. These modes provide
different levels of experimental evidence. Ideal simulation can be used to
verify the implementation and behaviour of an algorithm under perfect
conditions. Simulation with a noise model provides an estimate of how the
algorithm may behave when specific hardware errors are introduced. Physical
hardware experiments additionally include effects such as calibration
variation, device drift, and correlated errors that may not be completely
represented by a simulator. This distinction is used when classifying the
studies discussed in Section~\ref{sec:applications}.

\subsection{Execution: primitives and the IBM Quantum Compute Service} \label{sub:execution}

Quantum workloads can be executed through the primitives introduced in
Section~\ref{sub:primitives}. An algorithm submits a compiled circuit
together with its parameter values and, when using the \texttt{Estimator},
the observables for which expectation values are required. The V2 interfaces
organise these inputs using PUBs and return results in array form.

Different implementations of the primitive interfaces are available for
different execution environments. The SDK provides reference implementations
for local use, Aer provides implementations for simulation, and IBM's managed
service, renamed the IBM Quantum Compute Service as described in
Section~\ref{sub:twoseries}, provides access to physical quantum hardware.

Capabilities such as error mitigation, session and batch execution modes,
and job scheduling depend on the particular execution implementation being
used. They are therefore properties of the execution environment rather than
of the primitive interface itself. This distinction is relevant when assessing
published experiments because stating that a study used the \texttt{Estimator},
for example, does not fully identify the execution environment. The
implementation and mitigation settings are also needed if the experiment is to
be reproduced.

\subsection{The ecosystem} \label{sub:ecosystem}

The \textit{Qiskit} ecosystem also includes several separately versioned packages.
Many of these packages originated from the restructuring of Aqua described
in Section~\ref{sub:decomposition}.

The main domain packages include \textit{Qiskit} Nature, \textit{Qiskit} Optimization,
\textit{Qiskit} ML, and \textit{Qiskit} Finance. They provide problem
representations and algorithm implementations for chemistry and physics,
combinatorial optimization, quantum ML (QML), and financial modelling,
respectively. These packages are used by many of the application studies
considered in this review. \textit{Qiskit} Experiments provides tools for
characterization, calibration, and benchmarking that were previously
associated with Ignis, while \texttt{qiskit-ibm-runtime} provides the client
interface for IBM's managed execution service.

The SDK and the ecosystem packages are versioned independently. As a result,
reporting only the \textit{Qiskit} SDK version does not completely specify the
software environment used in an experiment involving one or more domain
packages. The packages also follow their own release schedules, and their
support for changes in the core SDK may therefore occur at different times.
These dependencies add to the compatibility considerations discussed in
Section~\ref{sec:evolution}.

IBM also provides the \textit{Qiskit} Functions Catalog, a collection of higher
level services developed by IBM and partner organizations that abstract parts
of the quantum software workflow. Circuit functions accept abstract circuits
and observables and manage transpilation, error suppression, error mitigation,
and execution, while application functions accept classical inputs for a
specific problem domain and return solutions. Introduced in 2024, the catalog
has since expanded across chemistry, optimization, differential equations, and
ML. The catalog remains in preview, with access available through selected IBM
Quantum service plans.
Since the application studies reviewed in this paper mainly use the SDK, Aer,
domain packages, and direct IBM hardware access, \textit{Qiskit} Functions are
treated here as part of the contemporary software environment rather than as a
separate application layer ~\cite{qiskitfunctions}.

\subsection{The contemporary workflow} \label{sub:workflow}

Figure~\ref{fig:modern-workflow} summarises how the different parts of the
current \textit{Qiskit} environment interact. A domain problem is first converted
into the required circuits and observables, either directly through the SDK
or with the help of a domain package. The circuits are then compiled using
the \texttt{Target} associated with the intended execution environment so
that they satisfy its available instructions and connectivity constraints.

The compiled circuits are submitted through a primitive interface. The
implementation selected by the user determines whether the workload runs
locally, on an Aer simulator with or without a noise model, or on physical
hardware through the IBM Quantum Compute Service. The resulting measurement
information or expectation values are then returned to the application. In
variational algorithms, these results are passed to a classical optimizer,
which updates the circuit parameters before the next execution.

This workflow differs from the earlier element model in two important ways.
Compilation and submission are now handled as separate parts of the
workflow, whereas older versions commonly combined them through the
\texttt{execute()} function. Error mitigation is also configured through
the execution environment rather than being represented as a separate
stage through Ignis.

\begin{figure}
\centering

\begin{tikzpicture}[
    font=\small,
    >=Latex,
    main/.style={
        draw,
        rounded corners=2pt,
        align=center,
        minimum height=0.9cm,
        text width=4.6cm,
        inner sep=5pt
    },
    exec/.style={
        draw,
        rounded corners=2pt,
        align=center,
        minimum height=1.0cm,
        text width=3.4cm,
        inner sep=5pt
    },
    side/.style={
        draw,
        rounded corners=2pt,
        align=center,
        minimum height=1.0cm,
        text width=3.5cm,
        inner sep=5pt
    },
    note/.style={
        draw,
        dashed,
        rounded corners=2pt,
        align=center,
        text width=3.3cm,
        font=\footnotesize,
        inner sep=4pt,
        fill=white
    },
    flow/.style={
        ->,
        thick
    },
    feedback/.style={
        ->,
        thick,
        dashed
    }
]

\node[main] (problem) {
    \textbf{Domain Problem}
};

\node[main, below=0.55cm of problem] (representation) {
    \textbf{Problem Representation}\\[2pt]
    \textit{Qiskit} Nature \textit{Qiskit} Optimization\\
    \textit{Qiskit} ML\\ \textit{Qiskit} Finance
};

\node[main, below=0.55cm of representation] (sdk) {
    \textbf{\textit{Qiskit} SDK}\\[2pt]
    \texttt{QuantumCircuit} + Observables
};

\node[main, below=0.55cm of sdk] (compilation) {
    \textbf{Compilation}\\[2pt]
    Pass Manager
};

\node[side, left=1.6cm of compilation] (target) {
    \textbf{\texttt{Target}}\\[2pt]
    Instructions, connectivity,\\
    errors and durations
};

\node[main, below=0.55cm of compilation] (primitive) {
    \textbf{Primitive Interface}\\[2pt]
    \texttt{Sampler} $\vert$ \texttt{Estimator}
};

\draw[flow] (problem) -- (representation);
\draw[flow] (representation) -- (sdk);
\draw[flow] (sdk) -- (compilation);
\draw[flow] (target) -- (compilation);
\draw[flow] (compilation) -- (primitive);

\node[exec, below=1.0cm of primitive, xshift=-5.0cm] (local) {
    \textbf{Local Reference}\\
    Implementation
};

\node[exec, below=1.0cm of primitive] (aer) {
    \textbf{Aer Simulator}\\
    Ideal or noise model
};

\node[exec, below=1.0cm of primitive, xshift=5.0cm] (ibm) {
    \textbf{IBM Quantum}\\
    \textbf{Compute Service}\\
    Physical hardware
};

\draw[flow] (primitive.south) -- ++(0,-0.35)
    -| (local.north);

\draw[flow] (primitive) -- (aer);

\draw[flow] (primitive.south) -- ++(0,-0.35)
    -| (ibm.north);

\node[main, below=2.0cm of aer] (results) {
    \textbf{Results}\\[2pt]
    Samples or Expectation Values
};

\draw[flow] (local.south) |- (results.west);
\draw[flow] (aer) -- (results);
\draw[flow] (ibm.south) |- (results.east);

\node[note, below=0.45cm of ibm] (settings) {
    Execution settings\\
    Error mitigation, sessions,\\
    batches and job scheduling
};

\draw[dashed] (ibm) -- (settings);

\node[main, below=0.6cm of results] (optimizer) {
    \textbf{Classical Optimiser}\\[2pt]
    Updates circuit parameters
};

\draw[flow] (results) -- (optimizer);

\draw[feedback]
    (optimizer.east)
    -- ++(6.0cm,0)
    |- node[pos=0.72,right,font=\footnotesize,align=center]
       {Updated\\parameters}
    (sdk.east);

\end{tikzpicture}

\caption{Contemporary \textit{Qiskit} execution workflow using \textit{Qiskit} SDK 2.5.
A domain problem is represented using the \textit{Qiskit} SDK or an associated domain package. Circuits and observables are compiled for an explicit \texttt{Target} before execution through the \texttt{Sampler} or \texttt{Estimator} primitive interface. The selected implementation determines whether the workload is executed locally, on an Aer simulator,
or on physical hardware through the IBM Quantum Compute Service. For variational algorithms, the returned results are processed by a classical optimizer and the updated parameters are supplied to the next circuit evaluation.}
\label{fig:modern-workflow}

\end{figure}

\section{Major Application Domains} \label{sec:applications}

The application literature considered in this review is organized into four areas: cryptography and cybersecurity, image and signal processing, climate science and energy applications, and finance and risk analysis. Within each area, the studies are examined in terms of the problem addressed, the quantum method used, the role of \textit{Qiskit} in the implementation, and the type of experimental evidence reported. This allows results obtained through ideal simulation, noisy simulation, and physical hardware to be considered separately.

These four areas were selected because each contains published studies in
which \textit{Qiskit} is used directly in the implementation rather than
mentioned only in passing. They also represent different classes of quantum
methods, including fixed circuit constructions, quantum kernels, variational
algorithms, and amplitude estimation. Across the four areas, the reviewed
studies include ideal simulation, simulation with noise models, and execution
on physical quantum processors. This provides a common basis for examining the
level of experimental evidence reported across different application problems.
The four areas are not intended to represent every field in which
\textit{Qiskit} has been applied. Other application areas also contain
\textit{Qiskit} based research, but they fall outside the scope of the present
review. Related applications are also developed using other quantum software
platforms, including \textit{PennyLane}, \textit{Cirq}, and \textit{CUDA-Q}, and some of
the studies discussed below use more than one framework. This review does not
compare these platforms and instead focuses on the role of \textit{Qiskit} in
the implementations examined here. The following subsections consider the four
selected areas in turn.

\subsection{Cryptography and Cybersecurity} \label{subsec:cryptography}

QC has important implications for cybersecurity because
algorithms such as Shor's algorithm can solve integer factorization and
discrete logarithm problems efficiently on a sufficiently capable quantum
computer~\cite{shor1999polynomial}. Quantum information also provides
cryptographic methods whose security depends on properties of quantum
states. The \textit{Qiskit} literature in this area includes implementations of
quantum key distribution, quantum signatures, authorization protocols,
quantum adaptations of hashing algorithms, and symmetric quantum
encryption.

\subsubsection{Quantum Key Distribution}

QKD allows two communicating parties to establish a shared secret key using quantum state preparation and measurement~\cite{renner2008security}. Disturbances introduced during transmission can
be detected through changes in the measured outcomes. The Bennett Brassard 1984 (BB84) protocol is one of the most widely studied examples and relies on the use of non-orthogonal quantum states together with the fact that an unknown quantum state cannot be copied perfectly~\cite{wootters1982single,shor2000simple}.

For each position \(i\), Alice selects a classical bit and a preparation basis. The state prepared by Alice can be written as

\begin{equation}
\lvert \psi_i\rangle =
H^{A_i}X^{a_i}\lvert 0\rangle .
\end{equation}

Here, \(i\) denotes the position of a transmitted bit, \(a_i\in\{0,1\}\)
is Alice's classical bit, and \(A_i\in\{0,1\}\) represents her basis
choice. \(A_i=0\) corresponds to the computational basis and \(A_i=1\)
corresponds to the diagonal basis. \(X\) denotes the Pauli \(X\) operator,
\(H\) denotes the Hadamard operator, and \(\lvert\psi_i\rangle\) is the
quantum state prepared for position \(i\).

Bob independently selects a measurement basis \(B_i\in\{0,1\}\). Before
measurement, he applies \(H^{B_i}\). When Alice and Bob select the same
basis, the preparation and measurement operations are consistent and Bob
recovers Alice's encoded bit in an ideal channel. Positions for which the
bases differ are discarded during key sifting.

A compact implementation using the current \textit{Qiskit} execution style is shown
below. Aer is imported from the separately installed
\texttt{qiskit\_aer} package, and the circuit is transpiled before
simulation.

\begin{lstlisting}[style=Pythonstyle]
import numpy as np

from qiskit import QuantumCircuit, transpile
from qiskit_aer import AerSimulator

n = 4
rng = np.random.default_rng()

# Alice chooses bits and preparation bases
alice_bits = rng.integers(0, 2, size=n)
alice_bases = rng.integers(0, 2, size=n)

qc = QuantumCircuit(n, n)

for i in range(n):
    if alice_bits[i] == 1:
        qc.x(i)
    if alice_bases[i] == 1:
        qc.h(i)

# Bob chooses measurement bases
bob_bases = rng.integers(0, 2, size=n)

for i in range(n):
    if bob_bases[i] == 1:
        qc.h(i)

qc.measure(range(n), range(n))

# Local simulation
backend = AerSimulator()
compiled = transpile(qc, backend)

result = backend.run(
    compiled,
    shots=1,
    memory=True
).result()

measured = result.get_memory()[0][::-1]
bob_bits = np.array([int(bit) for bit in measured])

# Retain positions with matching bases
keep = alice_bases == bob_bases

alice_key = alice_bits[keep].tolist()
bob_key = bob_bits[keep].tolist()
\end{lstlisting}

The quantum operations applied to one position of the protocol are shown in
Fig.~\ref{fig:bb84_protocol}. Since the exponents \(a_i\), \(A_i\), and
\(B_i\) take binary values, an exponent of zero represents the identity
operation and an exponent of one applies the corresponding gate.

\begin{figure}
\centering

\begin{quantikz}[column sep=1.3cm, row sep=0.7cm]
\lstick{$\ket{0}$}
    & \gate{X^{a_i}}
    & \gate{H^{A_i}}
    & \gate{H^{B_i}}
    & \meter{}
\end{quantikz}

\caption{Quantum operations for one position of the BB84 protocol.
Alice encodes the classical bit \(a_i\) and selects the preparation basis
using \(A_i\). Bob selects the measurement basis using \(B_i\). Positions
for which the two basis choices agree are retained during classical key
sifting.}
\label{fig:bb84_protocol}

\end{figure}

After Alice and Bob compare their basis choices, only positions for which
\(A_i=B_i\) are retained. Let \(\mathcal{I}\) denote this set of indices.
Errors in the retained key can then be quantified using the quantum bit
error rate,

\begin{equation}
\mathrm{QBER}
=
\frac{1}{|\mathcal{I}|}
\sum_{i\in\mathcal{I}}
\mathbf{1}[a_i\neq b_i].
\end{equation}

Here, \(|\mathcal{I}|\) is the number of retained positions, \(b_i\) is
Bob's measured classical bit at position \(i\), and
\(\mathbf{1}[\cdot]\) is the indicator function, which equals one when the
stated condition is true and zero otherwise. An increased Quantum bit error rate (QBER) can indicate
disturbance of the transmitted states, although practical implementations
must also account for errors introduced by the communication channel and
the hardware.

The circuit in Fig.~\ref{fig:bb84_protocol} represents the quantum state
preparation and measurement stages of BB84. A complete QKD system also
requires authenticated classical communication, parameter estimation,
error correction, and privacy amplification.

The literature reviewed in this area extends these principles to several
other security tasks. Mohanty \textit{et al.}
~\cite{mohanty2024experimentally} proposed an identity based quantum
signature protocol for applications such as secure email communication.
Their construction uses a quantum one time pad during signing and
verification and associates each signer with secret information shared
through a secret key generator. The protocol was evaluated using a \textit{Qiskit}
simulator, followed by a small implementation on physical quantum hardware.
The reported experiment demonstrates the circuit procedure for a small
instance, while the security properties of the complete protocol are
established analytically rather than through the hardware experiment alone.

Faleiro and Goul\~{a}o~\cite{faleiro2021device} studied device independent
quantum authorization using the Clauser Horne Shimony Holt (CHSH) game. The
authorization procedure is based on correlations obtained from different
measurement settings. The CHSH parameter can be expressed as

\begin{equation}
S =
E(A_0,B_0)
+
E(A_0,B_1)
+
E(A_1,B_0)
-
E(A_1,B_1),
\end{equation}

where \(S\) is the CHSH parameter, \(A_0\) and \(A_1\) are Alice's two
measurement settings, \(B_0\) and \(B_1\) are Bob's corresponding settings,
and \(E(A_x,B_y)\) denotes the correlation between the outcomes obtained for
a selected pair of settings. Local classical models satisfy
\(|S|\leq 2\), while quantum correlations can reach the Tsirelson bound
\(|S|\leq 2\sqrt{2}\). Values above the classical limit indicate
correlations that cannot be reproduced by a local hidden variable model.
The authors implemented the authorization procedure as a \textit{Qiskit} proof of
concept using simulation. The experiment verifies the circuit logic of the
protocol, although a complete deployment would also require independent
physical devices and a communication channel.

Das \textit{et al.}~\cite{das2023design} investigated \textit{Qiskit} circuit
implementations of Secure Hash Algorithm (SHA) 1, Message Digest Algorithm 5 (MD5), and SHA 256. Their circuits reproduce
operations used in the corresponding classical hash functions, including
logical transformations, addition, and modular arithmetic. The authors
reported experiments using \textit{Qiskit} simulation and IBM quantum hardware.
Their quantum implementations required considerably more execution time
than the corresponding classical implementations, and the measured times
also varied with the classical processor used in the experiments. The
study therefore demonstrates how conventional hashing operations can be
represented as quantum circuits, but it does not provide evidence of a
computational advantage over classical hashing.

Perepechaenko and Kuang~\cite{kuang2022quantum} considered symmetric
quantum encryption using the Quantum Permutation Pad and implemented the
method through \textit{Qiskit} on the IBM \texttt{ibmq\_manila} system. Plaintext was divided into two bit blocks,
encoded using two qubits, and transformed by permutation operators selected
from the symmetric group \(S_4\). This group contains \(4!=24\) possible
permutations, so selecting one permutation corresponds to approximately
\(\log_2(24)\approx4.58\) bits of information entropy. A pad containing
28 such permutations therefore corresponds to approximately 128 bits of
information entropy in the proposed construction.

The hardware experiment also reflects limitations of the available quantum
systems. Since a quantum communication channel was not available, the
encrypted qubits were measured and the resulting classical ciphertext was
transmitted before the decryption stage. Repeated executions were used to
reduce the effect of noisy hardware, and the authors described the
implementation as a simplified demonstration of the Quantum Permutation Pad
procedure. They also noted that the secret used in the experiment was not
itself a 128 bit key. The reported value of 128 bits therefore refers to the
information entropy associated with the permutation pad construction and
should not be interpreted as an experimental demonstration of 128 bit end
to end cryptographic security.

\subsubsection{Quantum Random Number Generation}

The measurement process used in QKD also provides a direct route to random
number generation, where the measurement outcomes themselves form the
random sequence. Quantum random number generators (QRNG) use the probabilistic
outcomes of quantum measurements as a source of entropy and are relevant
to applications such as cryptographic key generation, nonces, and other
security parameters~\cite{herrero2017quantum}. Their implementation on
programmable quantum computers provides a simple way to study this process,
although the distinction between simulation and physical hardware is
important when interpreting the resulting randomness.

For \(n\) qubits initially prepared in the state
\(\lvert 0\rangle^{\otimes n}\), applying the \(H\) operator introduced in the preceding discussion produces

\begin{equation}
\lvert \Psi_n\rangle =
H^{\otimes n}\lvert 0\rangle^{\otimes n}
=
\frac{1}{\sqrt{2^n}}
\sum_{x=0}^{2^n-1}\lvert x\rangle .
\end{equation}

Here, \(n\) is the number of qubits, \(x\) denotes one of the \(2^n\)
computational basis states, and \(\lvert\Psi_n\rangle\) is the resulting
uniform superposition. Under ideal measurement, every \(n\) bit outcome
therefore occurs with probability \(1/2^n\). For eight qubits, the measured
bit string can be interpreted as an integer from 0 to 255.

The following example implements this circuit using the current \textit{Qiskit}
execution style. Ten measurement outcomes are requested from a single
compiled circuit and converted to integer values. The example uses
\texttt{AerSimulator} so that it can be reproduced locally. It illustrates
the circuit and sampling procedure, but the simulator should not itself be
treated as a physical source of quantum entropy.

\begin{lstlisting}[style=Pythonstyle]
from qiskit import QuantumCircuit, transpile
from qiskit_aer import AerSimulator

n_bits = 8
n_numbers = 10

# Construct the QRNG circuit
qc = QuantumCircuit(n_bits, n_bits)

qc.h(range(n_bits))
qc.measure(range(n_bits), range(n_bits))

# Compile and execute locally
backend = AerSimulator()
compiled = transpile(qc, backend)

result = backend.run(
    compiled,
    shots=n_numbers,
    memory=True
).result()

# Retrieve individual measurement outcomes
bitstrings = result.get_memory(compiled)

# Convert the binary outcomes to integers
random_numbers = [
    int(bits, 2) for bits in bitstrings
]

print(random_numbers)
\end{lstlisting}

The corresponding circuit is shown in Fig.~\ref{fig:QRNGckt}. No ancillary
qubit is required for this basic construction because every measured qubit
contributes directly to the output string.

\begin{figure}
\centering

\begin{quantikz}[row sep=0.3cm, column sep=1.2cm]
\lstick{$\ket{0}_{q_0}$} & \gate{H} & \meter{} \\
\lstick{$\ket{0}_{q_1}$} & \gate{H} & \meter{} \\
\lstick{$\ket{0}_{q_2}$} & \gate{H} & \meter{} \\
\lstick{$\ket{0}_{q_3}$} & \gate{H} & \meter{} \\
\lstick{$\ket{0}_{q_4}$} & \gate{H} & \meter{} \\
\lstick{$\ket{0}_{q_5}$} & \gate{H} & \meter{} \\
\lstick{$\ket{0}_{q_6}$} & \gate{H} & \meter{} \\
\lstick{$\ket{0}_{q_7}$} & \gate{H} & \meter{}
\end{quantikz}

\caption{Eight qubit circuit illustrating the basic QRNG procedure. Each
qubit is prepared in an equal superposition and then measured in the
computational basis. Under the ideal quantum model, the resulting eight bit
strings are uniformly distributed over the integers from 0 to 255.
Physical implementations may exhibit bias and correlations caused by gate,
readout, and other hardware errors.}
\label{fig:QRNGckt}

\end{figure}

The basic circuit describes the source of the measurement outcomes but does
not by itself establish that an implementation is suitable for
cryptographic use. On physical hardware, preparation errors, gate errors,
readout bias, and temporal correlations can alter the output distribution.
Randomness extraction and statistical testing are therefore common parts
of practical QRNG implementations. Statistical tests can identify properties
of an observed sequence, but passing such tests alone does not establish
the physical origin or security of the entropy.

Germain \textit{et al.}~\cite{germain2022qubit} used \textit{Qiskit}'s reset
functionality to address the relation between the desired output length and
the number of available qubits. Their implementation constructs a circuit
in several phases and resets qubits so that they can be reused within the
same random number generation procedure. This allowed the output length to
exceed the number of qubits available on the selected IBM quantum system.
The authors generated 1000 binary numbers of 1024 bits and 40,000 binary
numbers of 25 bits and evaluated the resulting sequences using the National Institute of Standards and Technology (NIST)
statistical test suite. They reported strong results across the tests
performed. The main contribution of this study is the use of reset and
reuse to generate long random strings with a small quantum register rather
than an increase in the entropy generated by each individual measurement.

Kumar \textit{et al.}~\cite{kumar2022quantum} examined a 24 qubit
construction using \textit{Qiskit} 0.29.0. Their circuits contained odd numbers of
consecutive \(H\) gates, using one, three, five, seven, and nine gates
per qubit, and were evaluated with 65,536 shots using the
\texttt{ibmq\_qasm\_simulator}. The authors reported a min entropy value of
0.0007244 and a corresponding worst case entropy value of 0.999507, along
with restart analysis, autocorrelation analysis, and tests from NIST
SP 800 90B and SP 800 22. These results describe the statistical behaviour
of the generated sequences under the reported simulation procedure.
Because the experiment used the \textit{Qiskit} simulator rather than measurements
from a physical quantum processor, it does not by itself demonstrate a
hardware source of quantum entropy. This distinction is important when the
results are interpreted in a cryptographic setting.

A stronger treatment of hardware imperfections was presented by Li
\textit{et al.}~\cite{li2021quantum}, who implemented a source independent
QRNG on IBM superconducting quantum computers. Their method prepares the
state \(\lvert+\rangle\) using an \(R_Y(\pi/2)\) rotation and performs
measurements in the \(X\) and \(Z\) bases. Measurements in the \(X\) basis
are used to estimate errors in the preparation of the superposition state,
while measurements in the \(Z\) basis provide the raw random sequence.
Toeplitz matrix hashing is then applied to extract the final random bits.

The number of extractable bits in the presence of asymmetric readout errors
is bounded in their analysis by

\begin{equation}
K_{\mathrm{final}}
=
(1-r_1+r_0)
\left[
n_z-n_z h_2(e_z)
\right]
-t_e .
\end{equation}

Here, \(K_{\mathrm{final}}\) is the number of extracted random bits,
\(r_0\) is the probability of reading a prepared
\(\lvert0\rangle\) state as \(\lvert1\rangle\), and \(r_1\) is the
probability of reading \(\lvert1\rangle\) as \(\lvert0\rangle\).
The expression follows the convention \(r_0\leq r_1\) used in the
analysis. \(n_z\) is the number of raw outcomes obtained from measurements
in the \(Z\) basis, \(e_z\) is the estimated error in preparing the
superposition state, \(h_2(e_z)\) is its binary Shannon entropy, and
\(t_e\) determines the failure probability associated with randomness
extraction.

The experiment used \texttt{IBMQ\_5\_yorktown} and
\texttt{IBMQ\_lima}, both of which were five qubit devices, with qubit
\(Q_0\) selected for the reported measurements. The authors repeatedly
executed 8192 measurements in the \(Z\) basis and 251 in the \(X\) basis,
eventually collecting 819,200 raw \(Z\) basis outcomes and 25,100
\(X\) basis outcomes for each device. Their analysis yielded 741,006
extractable random bits for Yorktown and 621,729 for Lima, corresponding
to extraction ratios of 0.9045 and 0.7589. These values are extraction
ratios rather than bit rates. After Toeplitz hashing, 600,000 bit sequences
were evaluated using the nine NIST test items supported by the available
data length, with reported \(p\) values above 0.01. The measured
autocorrelation coefficients also remained below the reported three
standard deviation threshold of approximately 0.003873. This experiment
provides hardware evidence while explicitly accounting for preparation and
readout errors rather than assuming an ideal measurement process.

Orts \textit{et al.}~\cite{orts2023quantum} considered a different
problem: generating random integers in a specified interval
\([0,B)\) when \(B\) is not necessarily a power of two. A simple \(H\)
circuit naturally produces values over the complete range represented by
its qubits, so restricting the output interval without introducing unwanted
bias requires additional circuit operations. Their design combines a
reversible comparator with Grover amplitude amplification so that states
representing values inside the required interval receive greater
measurement probability.

The comparator was designed using temporary logical AND operations and,
for an \(N\) bit number, requires \(2N+1\) qubits with a reported
\(T\) count of \(4N\) and \(T\) depth of \(2N\). The complete random
number generator requires \(2N+2\) qubits. The implementation was developed
with \textit{Qiskit} 0.36.2. Comparator experiments used the matrix product state
simulator, while the random number generator was evaluated with the noisy
\texttt{ibmq\_qasm\_simulator}. For circuits covering three to eight bit
numbers, the authors reported that an appropriate choice of qubit count
could keep the probability of obtaining a value inside the requested
interval above 80 percent, with some interval configurations exceeding
90 or 95 percent. Their noise experiments also showed reduced uniformity,
so the study does not claim that the circuit removes the need for
randomness correction on noisy devices.

Sinai and In~\cite{sinai2024q} used a QRNG as part of their Quantum Random
Transaction Ordering Protocol for blockchain transactions. Their \textit{Qiskit}
circuit applies \(H\) gates and measurements to obtain random indices
that are used when shuffling transactions before block validation. Five
qubits, for example, represent values from 0 to 31, while their simulation
of 8192 transactions used thirteen qubits. The authors report that the
complete transaction randomization procedure required approximately
25 ms for 8192 transactions in their experimental environment. The QRNG
evaluation itself used IBM's quantum simulator because physical hardware
jobs could involve substantial queueing delays. The reported timing
therefore characterizes the simulated \textit{Qiskit} based transaction ordering
workflow and should not be interpreted as the latency of an operational
hardware QRNG or as direct evidence of cryptographic security against a
quantum adversary.

The studies in this area consequently provide different forms of evidence.
Hardware experiments such as those of Germain \textit{et al.} and Li
\textit{et al.} examine randomness obtained from measurements on physical
quantum systems, while the experiments of Kumar \textit{et al.}, Orts
\textit{et al.}, and Sinai and In rely primarily on simulation for the
reported \textit{Qiskit} workflows. The distinction affects what can be concluded
from each result, especially when the generated numbers are intended for
cryptographic use.

\subsection{Image and Signal Processing} \label{2.2}

Following the cryptography studies, the next group of applications concerns
the representation and analysis of image and signal data with quantum
circuits. Current work in this area includes quantum image representations,
quantum kernels, variational circuits, and hybrid learning models in which
classical feature extraction is combined with a smaller quantum component.
\textit{Qiskit} has been used both to construct these circuits and to evaluate them
through simulation or quantum hardware. The available studies remain
largely experimental, particularly because high resolution image data must
usually be reduced before it can be processed using present quantum
resources ~\cite{elaraby2022quantum}.

\subsubsection{Medical Imaging}

Medical imaging provides a practical setting for studying these methods
because image classification and representation involve large numerical
data while current quantum processors support relatively small circuits.
Most \textit{Qiskit} studies therefore do not place a complete medical image directly
on a quantum processor. Instead, they either study quantum representations
of image data or use a classical model to extract a small feature vector
that can be processed by a quantum circuit.

One of the established methods for representing an image as a quantum state
is the Flexible Representation of Quantum Images (FRQI), introduced by
Le \textit{et al.}~\cite{le2011flexible}. For a grayscale image of size
\(2^m \times 2^m\), FRQI uses \(2m\) qubits to represent the pixel
positions and one additional qubit to encode their intensities. The
resulting state is

\begin{equation}
\lvert I\rangle =
\frac{1}{2^m}
\sum_{j=0}^{2^{2m}-1}
\left(
\cos\theta_j\lvert 0\rangle_c
+
\sin\theta_j\lvert 1\rangle_c
\right)
\otimes
\lvert j\rangle_p .
\end{equation}

Here, \(\lvert I\rangle\) denotes the encoded image, \(m\) determines its
spatial dimensions, \(j\) identifies a pixel position, and
\(\lvert j\rangle_p\) is the corresponding state of the position register.
The subscript \(c\) denotes the intensity qubit, while
\(\theta_j\in[0,\pi/2]\) represents the grayscale intensity associated with
pixel \(j\). For an eight bit grayscale value \(g_j\in[0,255]\), a simple
linear mapping is
\(\theta_j=(\pi/2)(g_j/255)\). A value of
\(\theta_j=0\) therefore produces the intensity state
\(\lvert0\rangle_c\), while \(\theta_j=\pi/2\) produces
\(\lvert1\rangle_c\). Intermediate values produce superpositions whose
amplitudes encode intermediate grayscale levels.

A \(2\times2\) image requires two position qubits and one intensity qubit.
The following implementation uses four example grayscale values and the
current \textit{Qiskit} circuit interface. The position qubits are first placed in
an equal superposition. A multiple controlled \(R_y\) rotation is then
applied for each pixel address. Since \(R_y(2\theta_j)\lvert0\rangle
=
\cos\theta_j\lvert0\rangle+
\sin\theta_j\lvert1\rangle\), the rotation angle supplied to the circuit is \(2\theta_j\).

\begin{lstlisting}[style=Pythonstyle]
import numpy as np
from qiskit import QuantumCircuit

# Example 2 x 2 grayscale image
pixels = np.array(
    [0, 85, 170, 255],
    dtype=float
)

# Map grayscale values to FRQI angles
angles = (np.pi / 2) * (pixels / 255.0)

# q0 and q1: position register
# q2: intensity qubit
qc = QuantumCircuit(3)

position = [0, 1]
intensity = 2

# Equal superposition of the four pixel positions
qc.h(position)

# Encode each pixel intensity
for index, angle in enumerate(angles):

    bits = [
        (index >> bit) & 1
        for bit in range(2)
    ]

    # Convert controls on |0> to controls on |1>
    for qubit, value in zip(position, bits):
        if value == 0:
            qc.x(qubit)

    qc.mcry(
        2 * angle,
        position,
        intensity
    )

    # Restore the position register
    for qubit, value in zip(position, bits):
        if value == 0:
            qc.x(qubit)

qc.draw("mpl")
\end{lstlisting}

The corresponding logical circuit is shown in
Fig.~\ref{frqi_circuit}. Open controls indicate that a rotation is selected
when the corresponding position qubit is in state \(\lvert0\rangle\),
while filled controls correspond to \(\lvert1\rangle\). In the \textit{Qiskit} code,
the open controls are implemented by applying \(X\) gates before and after
the multiple controlled rotation.

\begin{figure}
\centering
\begin{quantikz}[column sep=0.75cm, row sep=0.55cm]
\lstick{$p_1:\ket{0}$}
    & \gate{H}
    & \octrl{2}
    & \octrl{2}
    & \ctrl{2}
    & \ctrl{2}
    & \qw \\
\lstick{$p_0:\ket{0}$}
    & \gate{H}
    & \octrl{1}
    & \ctrl{1}
    & \octrl{1}
    & \ctrl{1}
    & \qw \\
\lstick{$c:\ket{0}$}
    & \qw
    & \gate{R_y(2\theta_0)}
    & \gate{R_y(2\theta_1)}
    & \gate{R_y(2\theta_2)}
    & \gate{R_y(2\theta_3)}
    & \qw
\end{quantikz}

\caption{FRQI encoding circuit for a \(2\times2\) grayscale image.
The qubits \(p_1\) and \(p_0\) represent the four pixel positions, while
the qubit \(c\) stores the corresponding grayscale information through
controlled \(R_y\) rotations. Each rotation is conditioned on one of the
four position states.}
\label{frqi_circuit}
\end{figure}

The circuit prepares the FRQI state but does not measure it because
measurement is not required for the encoding stage itself. The use of only
three qubits for this example illustrates the compact state representation
provided by FRQI. This should not be interpreted as evidence that a complete
image processing task is automatically faster on a quantum computer.
Preparing all of the pixel dependent rotations and recovering information
from the final quantum state remain part of the computational cost
~\cite{le2011flexible,elaraby2022quantum}.

\textit{Qiskit} has also been used in medical image classification, where the quantum
circuit usually processes features that have already been extracted by a
classical network. Decoodt \textit{et al.}~\cite{decoodt2023hybrid}
studied cardiomegaly detection from chest radiographs using this approach.
They constructed a balanced dataset of 2436 posteroanterior chest
radiographs from the CheXpert repository and used a pretrained DenseNet121
network for feature extraction. A parameterized quantum circuit (PQC) then
replaced part of the classical classifier. Both \textit{Qiskit} and \textit{PennyLane}
implementations were examined.

For the \textit{Qiskit} models, the authors used
\texttt{CircuitQNN} and \texttt{TorchConnector} from \textit{Qiskit} ML to connect the quantum circuit with PyTorch. Their \textit{Qiskit}
experiments used four qubit circuits. After \(H\) initialization,
features were encoded through \(R_y\) rotations, followed by repeated
layers of controlled-not (CNOT) entanglement and parameterized \(R_y\) rotations. Training
and evaluation were performed using simulation rather than an IBM quantum
processor.

Across the classical quantum models, the study reported receiver operating characteristic (ROC) area under the curve (AUC) values as
high as 0.93 and accuracies as high as 0.87. These values were comparable
with the classical reference model rather than consistently higher. In the
ten fold comparison, the \textit{Qiskit} models had a lower AUC than the selected
classical reference, while most of the other predictive metrics did not
show a statistically significant difference. The study therefore supports
the feasibility of incorporating a small \textit{Qiskit} circuit into a medical
image classification pipeline, but it does not demonstrate a predictive
advantage over the classical model.

The authors also examined the regions used by the models when making a
prediction. In their blinded assessment, a cardiologist classified the
resulting Grad CAM++ maps as anatomically trustworthy when the highlighted
region covered the cardiac silhouette. Trustworthy maps were observed in
94\% of the evaluated images for one \textit{Qiskit} classical quantum configuration,
compared with 61\% for the classical reference. This result is relevant to
model interpretation, although the assessment was performed by a single
cardiologist and should not be treated as a clinical validation of the
classifier~\cite{decoodt2023hybrid}.

Reka \textit{et al.}~\cite{reka2024exploring} examined skin lesion
classification on the HAM10000 dataset using \textit{PennyLane} and \textit{Qiskit}. Their
study contained two separate quantum approaches. The quanvolutional network
was implemented with \textit{PennyLane} and processed \(2\times2\) image regions
using four qubits. The images used by this part of the experiment were
reduced to \(28\times28\) pixels because of the computational requirements
of the quantum simulation. Different combinations of \(R_x\), \(R_y\), and
\(R_z\) rotations and Pauli measurements were evaluated.

The \textit{Qiskit} part of the study used a Quantum Support Vector Classifier (QSVC).
Features were first extracted using MobileNet and then encoded using a
\texttt{ZZFeatureMap}. \textit{Qiskit}'s \texttt{FidelityQuantumKernel} was used to
construct the similarity matrix required by the classifier. For two
feature vectors \(x_i\) and \(x_j\), the fidelity kernel can be expressed as

\begin{equation}
K(x_i,x_j)
=
\left|
\langle\phi(x_i)\vert\phi(x_j)\rangle
\right|^2 .
\end{equation}

Here, \(x_i\) and \(x_j\) denote two classical feature vectors,
\(\lvert\phi(x_i)\rangle\) and
\(\lvert\phi(x_j)\rangle\) are the quantum states produced by the feature
map, and \(K(x_i,x_j)\) is their fidelity based similarity used by the
classifier.

The \textit{Qiskit} QSVC achieved a reported classification accuracy of 72.5\%.
For comparison, MobileNet, which produced the strongest result among the
classical pretrained networks in the study, achieved 73.42\%. The highest
accuracy reported in the paper, 82.86\%, came from the \textit{PennyLane}
quanvolutional model using \(R_y\) rotations and Pauli \(Z\) measurements.
The 82.86\% result should therefore not be attributed to the \textit{Qiskit}
implementation. The \textit{Qiskit} result instead shows that a fidelity based
quantum kernel produced classification performance close to the MobileNet
baseline in this particular experiment.

More recent work has evaluated a smaller quantum classifier on both
simulation and physical hardware. Guha \textit{et al.}
~\cite{guha2025resq} proposed ResQ for breast histopathology image
classification. Their architecture used a pretrained ResNet model to
extract features and a shallow one qubit variational quantum circuit as the
classifier. The model was evaluated on the BreakHis and Bioimaging
datasets. The authors reported accuracies of 98.49\% and 88.96\%,
respectively, compared with 97.36\% and 87.96\% for the corresponding
classical models.

The quantum circuit in ResQ was evaluated using the \textit{Qiskit} Aer simulator
and was also tested on the physical \texttt{ibm\_brisbane} processor under
hardware noise. The use of a single qubit kept the quantum part of the
architecture shallow, while most of the image processing remained in the
classical ResNet feature extractor. The reported improvements are therefore
results for this particular hybrid architecture and dataset rather than
evidence of a general quantum advantage in medical image classification.
The hardware experiment establishes that the small quantum component can be
executed on a current processor, but it does not demonstrate faster
processing of complete medical images.

Taken together, the medical imaging studies reviewed here use \textit{Qiskit} in
three different roles. FRQI provides a circuit level representation of
image intensity and position, \textit{Qiskit} ML allows variational
circuits to be integrated into classical neural networks, and quantum
kernels provide an alternative classifier for classically extracted
features. The experimental studies increasingly include physical hardware,
but their quantum components remain small and are normally preceded by
classical image resizing or feature extraction. Current results therefore
support \textit{Qiskit} as an experimental platform for medical image research,
while evidence for an end to end computational advantage over established
classical image processing remains limited.

\subsubsection{Telecommunications}

The use of compact quantum circuits for structured data also appears in
telecommunications, where the inputs are transmitted symbols, channel
conditions, and network variables rather than image features. Research in
this area has considered quantum Fourier transform (QFT) based transmission,
signal detection, and resource allocation. Most of the available evidence
comes from simulation or small quantum hardware experiments, so the results
are best interpreted as demonstrations of particular quantum formulations
rather than established improvements over current communication systems.

Quantum Orthogonal Frequency Division Multiplexing (OFDM) adapts ideas from
classical OFDM by using the QFT and its inverse
to process information represented by quantum states. One implementation
studied by Sabaawi \textit{et al.}~\cite{sabaawi2024exploiting} applies the
QFT before a simulated channel and the inverse QFT after the channel. A
simplified representation of this process is

\begin{equation}
\lvert\psi_{\mathrm{out}}\rangle
=
F^{\dagger}U_{\mathrm{ch}}(\theta)F
\lvert\psi_{\mathrm{in}}\rangle .
\end{equation}

Here, \(\lvert\psi_{\mathrm{in}}\rangle\) is the encoded input state,
\(F\) denotes the QFT, and \(F^{\dagger}\) denotes its inverse.
\(U_{\mathrm{ch}}(\theta)\) represents the channel model applied to the
qubits, with \(\theta\) specifying the rotation used to control the
strength of the disturbance. The state
\(\lvert\psi_{\mathrm{out}}\rangle\) is obtained immediately before
measurement. In the experiments considered below, the channel is represented
using single qubit \(R_x\) or \(R_y\) rotations. These rotations provide a
controlled model for studying circuit behaviour under disturbance and do
not represent every physical effect present in a wireless or quantum
communication channel.

The following \textit{Qiskit} example follows the QFT, channel, and inverse QFT
ordering used by Sabaawi \textit{et al.} A four bit message is first
encoded in the computational basis. The QFT is then applied across the
four qubits, followed by an \(R_x\) rotation on each qubit. An angle of
\(35^\circ\) is used because this value appears in one of the fixed angle
experiments reported in the study. The inverse QFT is applied before
measurement. The bit error rate is calculated as the fraction of measured
bits that differ from the transmitted message.

\begin{lstlisting}[style=Pythonstyle]
import numpy as np

from qiskit import QuantumCircuit, transpile
from qiskit.circuit.library import QFTGate
from qiskit_aer import AerSimulator

n_qubits = 4
message = "1010"
theta = np.deg2rad(35)
shots = 1024

qc = QuantumCircuit(n_qubits, n_qubits)

# Encode message in the computational basis
for q, bit in enumerate(reversed(message)):
    if bit == "1":
        qc.x(q)

# Apply QFT
qc.append(
    QFTGate(n_qubits),
    range(n_qubits)
)

# Apply a uniform channel rotation
for q in range(n_qubits):
    qc.rx(theta, q)

# Apply inverse QFT
qc.append(
    QFTGate(n_qubits).inverse(),
    range(n_qubits)
)

# Measure all qubits
qc.measure(
    range(n_qubits),
    range(n_qubits)
)

# Compile and simulate
backend = AerSimulator()
compiled = transpile(qc, backend)

result = backend.run(
    compiled,
    shots=shots
).result()

counts = result.get_counts(compiled)

# Calculate bit error rate
bit_errors = 0

for output, count in counts.items():
    bit_errors += count * sum(
        received != sent
        for received, sent
        in zip(output, message)
    )

ber = bit_errors / (n_qubits * shots)

print(counts)
print("BER:", ber)
\end{lstlisting}

The logical structure of the circuit is shown in
Fig.~\ref{fig:qft_telecom}. The states
\(\lvert s_0\rangle,\ldots,\lvert s_3\rangle\) denote the computational
basis states representing the four input bits. The QFT and inverse QFT act
on the complete register rather than on an individual qubit.

\begin{figure}
\centering
\begin{quantikz}[column sep=0.75cm, row sep=0.5cm]
\lstick{$\ket{s_0}$} & \gate[wires=4]{\mathrm{QFT}} & \gate{R_x(\theta)} & \gate[wires=4]{\mathrm{IQFT}} & \meter{} \\
\lstick{$\ket{s_1}$} & & \gate{R_x(\theta)} & & \meter{} \\
\lstick{$\ket{s_2}$} & & \gate{R_x(\theta)} & & \meter{} \\
\lstick{$\ket{s_3}$} & & \gate{R_x(\theta)} & & \meter{}
\end{quantikz}

\caption{Four qubit circuit illustrating a quantum OFDM model with a
rotation based channel. The QFT acts on the complete input register,
followed by an \(R_x(\theta)\) rotation on each qubit and an inverse QFT
before measurement. The rotation layer represents the controlled channel
disturbance used in the illustrative simulation.}
\label{fig:qft_telecom}
\end{figure}

The circuit in Fig.~\ref{fig:qft_telecom} is intended to show the processing
sequence rather than reproduce the complete communication model of a
particular experiment. The QFT distributes the information contained in the
input state across the register. The rotation layer then changes the state
before the inverse transformation and measurement. Repeating the experiment
over many shots provides a distribution of received bit strings from which
the bit error rate can be estimated. The same procedure can be repeated for
different rotation angles, input states, and numbers of qubits.

Almasaoodi \textit{et al.}~\cite{almasaoodi2024novel} developed a quantum
OFDM transmission model and evaluated it using \textit{Qiskit}. Their formulation
follows the ordering of classical OFDM more directly, with an inverse QFT at
the transmitter and a QFT at the receiver. The quantum channel was modelled
using \(R_x\) and \(R_y\) rotations, and the experiments were performed on a
four qubit register with 1024 shots for each configuration. The authors
compared the probability of recovering the transmitted state with a
reference quantum transmission circuit that did not contain the complete
OFDM transformation.

For one reported \(R_x\) experiment, the qubits were rotated by different
angles of \(30^\circ\), \(45^\circ\), \(60^\circ\), and \(90^\circ\).
The measured counts for the correct state were then compared with those of
the reference circuit. Further experiments applied a common rotation angle
to all four qubits and repeated the comparison over different values of the
angle. The quantum OFDM circuit generally produced a higher probability of
measuring the intended state than the reference circuit in these simulated
conditions. The result concerns the particular \textit{Qiskit} channel model used in
the paper and does not establish the performance of quantum OFDM over a
physical telecommunications channel.

Sabaawi \textit{et al.}~\cite{sabaawi2024exploiting} examined the same
general idea with a larger set of simulation conditions. Their experiments
used the \textit{Qiskit} Aer \texttt{qasm\_simulator} with registers containing
between two and eight qubits. Input bit strings were encoded as
computational basis states, transformed using the QFT, passed through a
channel containing \(R_x\) or \(R_y\) rotations, and processed with the
inverse QFT before measurement. The rotation angles included
\(15^\circ\), \(30^\circ\), \(45^\circ\), and \(60^\circ\), together
with randomly sampled values. Each experiment used 1024 shots.

The authors calculated the bit error rate across the possible input states
and compared the quantum OFDM circuit with a reference circuit that omitted
the QFT and inverse QFT stages. With randomly selected rotation angles and
five qubits, the reported average bit error rate was 1.40\% for the quantum
OFDM model and 8.10\% for the reference model. In another experiment,
\(\theta\) was fixed at \(35^\circ\) while the number of qubits was varied
from two to eight. The reported bit error rate of the quantum OFDM model
decreased from 2.801\% at two qubits to 0.8\% at eight qubits, whereas the
reference results remained close to 9\% over much of the tested range.
These values were obtained from \textit{Qiskit} simulations with rotation based
channel models. They should therefore be treated as comparisons between the
two simulated circuit constructions rather than measurements of an
operational OFDM communication link.

A different telecommunications problem was studied by Urgelles
\textit{et al.}~\cite{urgelles2024quantum}, who considered maximum
likelihood detection in Multiple Input Multiple Output (MIMO) Non Orthogonal
Multiple Access systems. The detection problem was converted to Quadratic
Unconstrained Binary Optimization and solved using the Quantum Approximate
Optimization Algorithm implemented through \textit{Qiskit} and IBM Quantum. The
binary optimization problem can be written in the form

\begin{equation}
C(\mathbf{x})
=
\mathbf{x}^{T}Q\mathbf{x}
+
\mathbf{L}^{T}\mathbf{x},
\end{equation}

where \(C(\mathbf{x})\) is the objective function,
\(\mathbf{x}\) is the vector of binary decision variables,
\(Q\) contains the quadratic coefficients, and
\(\mathbf{L}\) contains the linear coefficients. The resulting cost is
encoded into the quantum approximate optimization algorithm (QAOA) problem Hamiltonian, whose low energy states
correspond to candidate solutions of the detection problem.

The authors first considered a \(2\times2\) MIMO non orthogonal multiple access (NOMA) system using binary
phase shift keying. Classical maximum likelihood results calculated in
MATLAB were compared with QAOA results obtained through IBM Quantum. The
quantum circuit was sampled with 1024 shots. For experiments involving
10,000 transmitted bits, the classical and quantum bit error rate curves
were reported as being close for the two user case, with the bit error
rate decreasing as transmitted power increased.

The hardware experiment also provides information about execution cost.
The reported cloud job required approximately 12 hours from submission to
completion, but this interval included transpilation, validation, queue
time, and execution. The actual backend running time reported for the
circuit was 4.3 seconds. These quantities should not be treated as a direct
runtime comparison between the quantum and classical detection algorithms.
The study also considered larger MIMO configurations, where hardware noise
caused greater differences between the quantum and classical results. The
experiments therefore demonstrate a \textit{Qiskit} implementation of the quadratic unconstrained binary optimization (QUBO) and
QAOA formulation on current hardware while also showing the limitations
introduced by device noise and cloud execution.

Pathak \textit{et al.}~\cite{10628250} considered resource allocation in
5G networks using a Variational Quantum Regressor (VQR). Rather than formulating
symbol detection as a discrete optimization problem, their approach used a
parameterized quantum model to predict resource allocation values from
network related inputs. The reported experiments obtained a mean squared
error of 0.0081 for the VQR configuration considered in the study. This
value describes prediction error for that experimental setting and should
not be interpreted as a direct measurement of network latency, throughput,
or spectral efficiency.

The resource allocation study was simulation based and did not deploy the
model within an operational 5G radio access network. Its contribution is
therefore the formulation and evaluation of a variational quantum regression
model for a telecommunications resource allocation task. Further comparison
on common datasets and against tuned classical regression models would be
needed to determine whether the quantum model provides a consistent
advantage as the number of network variables increases.

Across the telecommunications studies considered here, \textit{Qiskit}
serves several different experimental roles. The quantum OFDM studies use
\textit{Qiskit} simulation to evaluate QFT based transmission circuits under
controlled rotation models. The MIMO NOMA study uses \textit{Qiskit} to formulate
and execute QAOA on IBM quantum systems, while the resource allocation
study applies a variational quantum model to network prediction. The
experiments cover relevant communication tasks, but they remain limited in
scale and rely heavily on simulation or small quantum circuits. The current
evidence therefore supports the use of \textit{Qiskit} for studying quantum
telecommunications algorithms, while an end to end performance advantage
in deployed 5G or 6G networks has not yet been established by these
experiments.

\subsection{Climate Science and Energy Applications} \label{2.3}

The communication applications discussed above use compact quantum
representations of signals and network variables. Climate and energy
research uses similar small quantum models for environmental data analysis,
surrogate modelling, and optimization. Direct \textit{Qiskit} studies in climate
science are currently less numerous than those in the preceding application
areas. Recent work has considered QML for Earth
observation data and quantum surrogate models for environmental simulation,
while broader proposals for quantum climate modelling remain largely
prospective.

Schwabe \textit{et al.}~\cite{schwabe2025} identify several areas in which QC could eventually contribute to climate models, including the solution of differential equations, representation of unresolved physical processes, model parameter tuning, and analysis of model output. Their study is a position paper rather than a \textit{Qiskit} implementation and is therefore used here to define the broader research context. The direct
\textit{Qiskit} evidence considered below concerns considerably smaller problems that can be evaluated with current simulators and quantum processors.

\subsubsection{Climate Modeling and Simulation}

Current \textit{Qiskit} applications in climate modelling operate on reduced
representations of environmental data rather than complete Earth system
models. This reduction is necessary because climate datasets contain many
spatial, temporal, and physical variables, whereas current quantum learning
experiments generally use only a small number of features and qubits. The
available studies therefore examine whether quantum circuits can contribute
to selected classification or surrogate modelling tasks within a larger
classical workflow.

Munasinghe \textit{et al.}~\cite{munasinghe2024qml_climate} conducted one
of the more direct evaluations of \textit{Qiskit} on climate and weather data. They
used Earth observation data obtained through NASA Giovanni for the state of
Texas. Total surface precipitation and surface soil wetness were selected
as the two input variables, while soil temperature was converted into a
binary target. Values above 295 K were labelled as warm and values below
295 K as cool. The resulting dataset contained 261 observations, with
189 samples in the cool class and 72 in the warm class. The authors
therefore evaluated the models on an imbalanced binary classification
problem rather than on a complete climate forecasting task.

Their quantum experiments used \textit{Qiskit} ML 0.7.2 and considered
both a QSVC and a Variational Quantum
Classifier. For the QSVC, the two environmental variables were encoded
using a two qubit \texttt{ZZFeatureMap} with linear entanglement and two
repetitions. State fidelity was then used to construct the kernel supplied
to the classifier. The fidelity kernel itself follows the same formulation
described earlier for quantum kernel classification, so it is not repeated
here.

The original study used the \textit{Qiskit} interfaces available in 2024. The
following code preserves its reported two feature map, linear entanglement,
two repetitions, and fidelity kernel while using the current \textit{Qiskit}
ML interface. The input array supplied to the function should
contain normalized precipitation and soil wetness measurements, with one
row for each observation.

\begin{lstlisting}[style=Pythonstyle]
import numpy as np

from qiskit.circuit.library import zz_feature_map
from qiskit.primitives import StatevectorSampler

from qiskit_machine_learning.algorithms import QSVC
from qiskit_machine_learning.kernels import (
    FidelityQuantumKernel
)
from qiskit_machine_learning.state_fidelities import (
    ComputeUncompute
)

def train_climate_qsvc(X_train, y_train):
    X_train = np.asarray(
        X_train,
        dtype=float
    )
    y_train = np.asarray(
        y_train,
        dtype=int
    )

    if X_train.ndim != 2 or X_train.shape[1] != 2:
        raise ValueError(
            "X_train must contain two features."
        )

    # Two feature configuration used in the study
    feature_map = zz_feature_map(
        feature_dimension=2,
        reps=2,
        entanglement="linear"
    )

    # Current reference primitive
    sampler = StatevectorSampler()

    fidelity = ComputeUncompute(
        sampler=sampler
    )

    quantum_kernel = FidelityQuantumKernel(
        feature_map=feature_map,
        fidelity=fidelity
    )

    model = QSVC(
        quantum_kernel=quantum_kernel
    )

    model.fit(
        X_train,
        y_train
    )

    return model
\end{lstlisting}

One repetition of the corresponding two qubit feature map is shown in
Fig.~\ref{fig:climate_zzfeature}. Let \(x_0\) denote the normalized total
surface precipitation value and \(x_1\) the normalized surface soil wetness
value for one observation. The phase gate \(P\) first encodes the
individual variables. The controlled operation introduces a phase depending
on both variables and therefore represents their interaction. The complete
configuration used in the code applies this block twice.

\begin{figure}
\centering
\begin{quantikz}[column sep=0.65cm, row sep=0.5cm]
\lstick{$\ket{0}_{q_0}$} & \gate{H} & \gate{P(2x_0)} & \ctrl{1} & \qw & \ctrl{1} & \qw \\
\lstick{$\ket{0}_{q_1}$} & \gate{H} & \gate{P(2x_1)} & \targ{} & \gate{P(2(\pi-x_0)(\pi-x_1))} & \targ{} & \qw
\end{quantikz}

\caption{One repetition of the two qubit ZZ feature map used to encode the
two environmental variables. The individual phase gates encode
\(x_0\) and \(x_1\), while the controlled phase operation represents their
joint contribution. The \textit{Qiskit} configuration used in the corresponding
classifier contains two repetitions of this block.}
\label{fig:climate_zzfeature}
\end{figure}

The feature map in Fig.~\ref{fig:climate_zzfeature} represents the data
encoding stage rather than the complete QSVC algorithm. For each climate
observation, the two classical variables determine the phase rotations in
the circuit. The fidelity between states prepared from different
observations is then evaluated to construct the kernel matrix used by the
classical support vector classifier. This division between quantum state
encoding and classical classification is important when interpreting the
computational role of the quantum circuit.

Munasinghe \textit{et al.} evaluated the QSVC and variational quantum classifier (VQC) using a \textit{Qiskit}
simulator and compared them with classical SVMs and a
Gaussian Naive Bayes classifier. On the 53 sample test set, the simulated
QSVC obtained an accuracy of 0.75 and a macro F1 score of 0.50. Its recall
for the warm class was only 0.07. The simulated VQC obtained an accuracy of
0.72 and a macro F1 score of 0.47, with the same 0.07 recall for the warm
class. These results show that overall accuracy alone gives an incomplete
view of performance because the dataset contains substantially more cool
than warm observations.

The authors also executed the VQC on a physical 127 qubit IBM Eagle
processor installed at Rensselaer Polytechnic Institute. Although the
processor contained 127 qubits, the classification problem itself was based
on the reduced two feature representation. The hardware experiment obtained
an overall accuracy of 0.74. For the warm class, however, precision, recall,
and F1 score were all zero. The classifier therefore failed to identify any
of the 14 warm observations in the test set. The reported accuracy mainly
reflects correct predictions for the majority cool class and should not be
interpreted as successful classification of both classes.

The classical comparisons provide an important reference for these results.
The linear and polynomial SVMs each obtained an accuracy
of 0.81 on the same test set. The linear SVM also achieved a macro F1 score
of 0.71, compared with 0.50 for the simulated QSVC and 0.42 for the
hardware VQC. The experiment is valuable because it evaluates a \textit{Qiskit}
model using real Earth observation data and physical quantum hardware, but
the reported results do not show a predictive advantage over the classical
models. The class imbalance and the use of only two environmental variables
also limit the conclusions that can be drawn about larger climate datasets.

A different form of environmental modelling was studied by Maheshwari
\textit{et al.}~\cite{maheshwari2025qcnn}, who considered surrogate
prediction of groundwater heat plumes produced by geothermal heat pumps in
Munich. Numerical groundwater simulations can provide detailed temperature
fields but must be recomputed when the physical conditions or heat pump
configuration changes. The authors investigated whether a Quantum
Convolutional Neural Network could learn a reduced representation of these
simulations and provide predictions without repeating the complete
numerical calculation.

The original spatial simulation data were reduced before being supplied to
the quantum model. The quantum convolutional neural network (QCNN) used PQCs for data
encoding, convolution, and pooling, followed by classical processing of
the measured quantum features. This reduction is central to the experiment:
the quantum circuit does not reproduce the full spatial groundwater
simulation. It learns a compact surrogate representation derived from the
classical simulation data.

An expanded study by the same authors
~\cite{maheshwari2026qcnn} reports experiments using \textit{Qiskit} 2.2.3 and
\textit{Qiskit} Runtime 0.45. The QCNN was evaluated with an ideal statevector
simulator, a simulator containing a hardware noise model, and the physical
127 qubit IBM Kyiv processor. A further hardware experiment applied error
mitigation to the same processor. Model quality was assessed using mean
squared error for the predicted heat plume quantities. Since mean squared
error has already been used as a performance measure in the preceding
applications, its standard definition is not repeated here.

Execution on the unmitigated Kyiv processor produced considerably larger
errors than the ideal and noisy simulations. The authors then applied a
mitigation configuration that included dynamical decoupling, readout
mitigation, and zero noise extrapolation together with additional mitigation
methods. The mitigated hardware experiment reduced the training and testing
loss relative to the unmitigated hardware run. It still did not reach the
performance obtained with the ideal simulator or the classical neural
network.

The classical model remained the most accurate predictor across the reported
configurations. This result is relevant when assessing the role of \textit{Qiskit}
in the study. The physical hardware experiment demonstrates that the QCNN
workflow can be executed on a current quantum processor and that mitigation
can reduce part of the error introduced by hardware noise. It does not show
that the quantum surrogate is more accurate or computationally faster than
the classical surrogate. The environmental simulation also undergoes
substantial dimensional reduction before quantum processing, so the
experiment does not constitute a quantum simulation of the complete
groundwater temperature field.

The climate studies discussed here consequently operate at a much smaller
scale than a full Earth system model. Munasinghe \textit{et al.} reduce
Earth observation data to two variables for classification, while
Maheshwari \textit{et al.} reduce a spatial groundwater simulation to a
compact surrogate learning problem. Both studies provide direct evidence
of \textit{Qiskit} being used with environmental data, and both include experiments
beyond ideal simulation. Their results also show the present limitations:
class imbalance, dimensional reduction, hardware noise, and stronger
classical baselines remain important factors. Among the studies reviewed
here, there is currently no \textit{Qiskit} implementation of a complete climate
model and no demonstrated computational advantage over established
classical climate or surrogate modelling methods.

\subsubsection{Energy Optimization and Smart Grids}

The environmental modelling studies above focus on prediction and surrogate
models, while energy systems also require decisions about which resources
should operate, how much power they should supply, and whether the grid
remains stable after a disturbance. Quantum optimization has therefore been
studied for problems such as unit commitment, while QML
has been applied to grid stability assessment. The \textit{Qiskit} literature in this
area consists mainly of small optimization instances, simulation studies,
and selected experiments on IBM quantum hardware.

Unit commitment is a standard power system optimization problem in which a
set of generating units must satisfy electricity demand at minimum operating
cost. The decision contains both discrete and continuous variables because a
generator can be switched on or off while its output, when active, can vary
within operating limits. A simplified single period formulation used in the
\textit{Qiskit} studies of Koretsky \textit{et al.}
~\cite{koretsky2021adapting} and Mahroo and Kargarian
~\cite{mahroo2022hybrid} is

\begin{equation}
\begin{aligned}
\min_{\{y_i,p_i\}}
\quad &
\sum_{i=1}^{N}
\left(
A_i y_i + B_i p_i + C_i p_i^2
\right) \\ 
\text{subject to}
\quad &
\sum_{i=1}^{N}p_i = L, \\
&
P_i^{\min}y_i
\leq p_i
\leq
P_i^{\max}y_i,
\qquad
y_i\in\{0,1\}.
\end{aligned}
\end{equation}

Here, \(N\) is the number of generating units, \(y_i\) specifies whether
unit \(i\) is committed, and \(p_i\) is its generated power.
\(A_i\), \(B_i\), and \(C_i\) are the coefficients of its operating cost.
The quantities \(P_i^{\min}\) and \(P_i^{\max}\) are the minimum and maximum
generation limits of the unit, while \(L\) is the required system load.
More detailed unit commitment models can also include ramp limits, reserve
requirements, minimum operating times, network constraints, and multiple
time periods. The studies considered here use reduced formulations so that
the binary component can be handled with a quantum optimization algorithm.

The binary decisions can be converted to a QUBO problem and represented by an Ising cost Hamiltonian. QAOA
then prepares a distribution over candidate commitment decisions and uses a
classical optimizer to adjust the circuit parameters
~\cite{farhi2014}. The following current \textit{Qiskit} example illustrates this
discrete part of the workflow. It uses four candidate generating units with
fixed illustrative outputs and costs. The example is intentionally smaller
than a complete unit commitment model because the purpose is to show how the
binary commitment stage is represented and solved using the present \textit{Qiskit}
Optimization interface.

\begin{lstlisting}[style=Pythonstyle]
from qiskit.primitives import StatevectorSampler

from qiskit_optimization import QuadraticProgram
from qiskit_optimization.algorithms import (
    MinimumEigenOptimizer
)
from qiskit_optimization.minimum_eigensolvers import (
    QAOA,
    NumPyMinimumEigensolver
)
from qiskit_optimization.optimizers import COBYLA

# Illustrative generator outputs and fixed costs
generation = [40, 55, 70, 85]
cost = [5, 7, 8, 12]
load = 125

# Binary commitment problem
qp = QuadraticProgram(
    name="unit_commitment_demo"
)

for i in range(4):
    qp.binary_var(
        name=f"y{i}"
    )

# Minimize total commitment cost
qp.minimize(
    linear={
        f"y{i}": cost[i]
        for i in range(4)
    }
)

# Generation must meet the required load
qp.linear_constraint(
    linear={
        f"y{i}": generation[i]
        for i in range(4)
    },
    sense="==",
    rhs=load,
    name="load_balance"
)

# QAOA using the current sampler interface
sampler = StatevectorSampler(
    seed=123
)

qaoa = QAOA(
    sampler=sampler,
    optimizer=COBYLA(
        maxiter=100
    ),
    reps=1,
    initial_point=[0.5, 0.5]
)

qaoa_solver = MinimumEigenOptimizer(
    qaoa,
    penalty=100.0
)

qaoa_result = qaoa_solver.solve(qp)

# Exact solution for comparison
exact_solver = MinimumEigenOptimizer(
    NumPyMinimumEigensolver()
)

exact_result = exact_solver.solve(qp)

print("QAOA result")
print(qaoa_result.prettyprint())

print("Exact result")
print(exact_result.prettyprint())
\end{lstlisting}

The equality constraint is converted internally into a penalty term when the
problem is mapped to QUBO form. Each binary variable is then represented by
one qubit. The exact solver is included only as a reference for this small
example so that the QAOA result can be checked against the optimum. QAOA is
an approximate method, and a shallow circuit is not guaranteed to return
the optimum on every problem instance.

A single QAOA layer for four commitment variables is shown in
Fig.~\ref{fig:qaoa_energy}. The cost operation acts on the complete
register, after which the mixer applies an \(X\) rotation to every qubit.

\begin{figure}
\centering
\begin{quantikz}[column sep=0.7cm, row sep=0.45cm]
\lstick{$\ket{0}_{y_0}$} & \gate{H} & \gate[wires=4]{U_C(\gamma)} & \gate{R_x(2\beta)} & \meter{} \\
\lstick{$\ket{0}_{y_1}$} & \gate{H} & & \gate{R_x(2\beta)} & \meter{} \\
\lstick{$\ket{0}_{y_2}$} & \gate{H} & & \gate{R_x(2\beta)} & \meter{} \\
\lstick{$\ket{0}_{y_3}$} & \gate{H} & & \gate{R_x(2\beta)} & \meter{}
\end{quantikz}

\caption{One QAOA layer for four binary generator commitment variables.
\(H\) gates prepare a superposition of the possible commitment states.
The cost operation \(U_C(\gamma)\) encodes the QUBO objective, and the
\(R_x(2\beta)\) rotations form the mixer before measurement.}
\label{fig:qaoa_energy}
\end{figure}

In Fig.~\ref{fig:qaoa_energy}, \(\gamma\) controls the phase applied by the
cost Hamiltonian and \(\beta\) controls the mixing rotation. The QAOA depth
determines how many cost and mixer layers are repeated. The circuit shown
contains one layer because it is intended to illustrate the structure of
the algorithm. The classical optimizer updates \(\gamma\) and \(\beta\)
using the sampled objective values until its stopping condition is reached.

Koretsky \textit{et al.}~\cite{koretsky2021adapting} developed a \textit{Qiskit}
implementation specifically for unit commitment. Their method separates the
two types of decisions in the problem. QAOA handles the binary variables
that indicate which generating units are operating, while a classical
optimizer determines their continuous power outputs. This avoids encoding
every possible power level as an additional set of qubits. The QAOA circuit
is simulated with \textit{Qiskit} during each iteration of the classical optimization
procedure.

The authors compared this formulation with classical approaches and examined
how the classical solution methods scale as the number of generating units
increases. Their conclusion requires careful interpretation. The study
reports that the classical solvers considered are effective for the
simulated unit commitment instances containing fewer than approximately
400 generating units. At larger problem sizes, the selected classical
methods either show rapidly increasing runtime or require coarser
approximations. The paper identifies systems containing several hundred
generating units as a possible scale at which future quantum methods would
need to become competitive.

The study does not demonstrate a \textit{Qiskit} QAOA solution of a unit commitment
problem containing more than 400 generating units. It therefore does not
establish quantum advantage at that scale. The authors themselves note that
quantum error correction may be required before quantum hardware can address
problems of the required size. Their result is more appropriately viewed as
an estimate of the problem scale at which a useful quantum implementation
would need to compete with established classical methods.

Mahroo and Kargarian~\cite{mahroo2022hybrid} considered the same unit
commitment problem using a different decomposition. Their method divides the
optimization into three blocks coordinated through the Alternating Direction
Method of Multipliers. Two of the blocks contain continuous variables and
are solved classically, while the block containing the binary commitment
variables is written as a QUBO and solved with QAOA.

The authors implemented the QAOA component in \textit{Qiskit} and used the IBM Q
environment for simulation. Their main case study contained ten generating
units. QAOA was configured with a circuit depth of two and a maximum of
100 optimization iterations. The quantum block was repeatedly coordinated
with the two classical blocks until the residual used by the decomposition
procedure satisfied its convergence condition.

For the tested load levels, the complete procedure was compared with a
centralized classical formulation. The authors report that the coordinated
quantum and classical method reproduced the commitment decisions obtained
by the classical reference when the QAOA parameters were updated during the
iterations. They also examined a four unit case with a 50 MW load and
reported that updating the variational parameters increased the probability
assigned to the desired commitment bit string.

These experiments establish that a \textit{Qiskit} QAOA block can be incorporated
into a decomposition method for a small unit commitment problem. The study
uses quantum simulation, however, and does not provide a runtime advantage
over a classical unit commitment solver. Its ten binary commitment variables
also represent a much smaller problem than the large utility scheduling
instances for which computational scaling becomes a practical concern.

Energy system applications of \textit{Qiskit} are not limited to dispatch decisions.
Zhou and Zhang~\cite{zhou2021noise} studied transient stability assessment,
which determines whether a power system can remain stable after a large
disturbance. Their quantum transient stability assessment method uses a
PQC to map system measurements into a quantum
feature space and then classify the operating state as stable or unstable.
The input features were generated from power system simulations and included
quantities such as generator power, rotor angle, rotor speed, transmission
line power flow, and bus voltage information.

The implementation used both \textit{Qiskit} and \textit{PennyLane}. Experiments were
performed with the 32 qubit \path{ibmq_qasm_simulator} and the real 20 qubit \texttt{ibmq\_boeblingen} processor. The authors considered a
single machine infinite bus system, a two area power system, and a larger
Northeast Power Coordinating Council (NPCC) system. The physical processor was used to examine how the trained
quantum classifier behaved in the presence of hardware noise.

For the test sets, the reported hardware accuracies were approximately
97.7\% for the single machine system, 99.1\% for the two area system, and
94.5\% for the NPCC case. The corresponding simulator results were about
98.1\%, 99.3\%, and 95.4\%. The small differences between the simulator and
hardware results in these experiments indicate that the particular shallow
circuits used in the study retained most of their classification performance
on that device.

The reported values should be interpreted within the scale and structure of
the experiments. The quantum circuit did not perform a complete dynamic
simulation of the power grid. Classical power system simulations first
generated the transient stability data, after which the quantum model
performed classification. The hardware experiment therefore demonstrates a
\textit{Qiskit} based quantum learning component within a larger power system
analysis workflow rather than a replacement for the underlying transient
simulation.

The energy studies reviewed here show two main uses of \textit{Qiskit}. Unit
commitment studies use QAOA for binary operating decisions while retaining
classical optimization for continuous power variables. Stability studies use
PQCs to classify states generated from classical
power system simulations. Both approaches provide concrete \textit{Qiskit}
implementations, including one study with physical IBM hardware, but the
reported problem sizes remain limited. The available results demonstrate
feasibility of these workflows and provide information about their behaviour
under simulation and hardware noise. They do not yet establish a
computational advantage over established classical optimization or power
system analysis methods.

\subsection{Finance and Risk Analysis} \label{2.4}

The optimization methods considered for energy systems also appear in
finance, where decisions are made under constraints involving expected
return, uncertainty, capital, and risk. \textit{Qiskit} has been used in this area
for portfolio selection, financial arbitrage, linear system formulations,
and risk estimation. The studies considered here include simulation,
historical backtesting, and experiments on physical quantum processors.
Their results need to be interpreted separately because these forms of
evaluation answer different questions about algorithm quality and practical
execution.

\subsubsection{Portfolio Optimization}

Portfolio optimization determines how capital should be allocated across a
set of assets according to an investment objective. In the classical
Markowitz formulation, expected return is considered together with the
variance of portfolio returns~\cite{markowitz2000mean}. The continuous
mean variance problem can be solved efficiently by established classical
optimization methods. The computational difficulty increases when the
problem contains discrete asset selection, integer holdings, cardinality
limits, transaction rules, or other constraints that introduce a
combinatorial search space.

\textit{Qiskit} Finance provides a binary portfolio formulation in which one qubit
is associated with each candidate asset~\cite{QPO}. For \(n\) assets, the
formulation can be written as

\begin{equation}
\begin{aligned}
\min_{\mathbf{x}\in\{0,1\}^{n}}
\quad &
q\,\mathbf{x}^{T}\Sigma\mathbf{x}
-
\boldsymbol{\mu}^{T}\mathbf{x}
\\
\text{subject to}
\quad &
\mathbf{1}^{T}\mathbf{x}=B .
\end{aligned}
\end{equation}

Here, \(\mathbf{x}\) is the binary asset selection vector, with
\(x_i=1\) when asset \(i\) is selected and \(x_i=0\) otherwise.
The vector \(\boldsymbol{\mu}\) contains the expected returns of the
assets, while \(\Sigma\) is their covariance matrix. The parameter
\(q>0\) determines the weight assigned to portfolio risk, \(B\) is the
number of assets to be selected, and \(\mathbf{1}\) is a vector whose
entries are all one. The quadratic term measures portfolio variance and
the linear term rewards expected return.

For quantum optimization, the budget constraint can be incorporated through
a penalty term and the resulting binary problem converted to QUBO form.
Using the correspondence \(x_i=(1-Z_i)/2\), where \(Z_i\) is the Pauli
\(Z\) operator acting on qubit \(i\), the QUBO can then be represented by
an Ising cost Hamiltonian. QAOA alternates evolution under this cost
Hamiltonian with a mixer that changes the probability assigned to different
asset selections. A classical optimizer updates the circuit parameters
using the objective values obtained from repeated circuit evaluations.

The following example uses three assets and selects exactly two of them.
The expected returns and covariance matrix are illustrative values rather
than market data. The implementation follows the current \textit{Qiskit} Optimization
interface and uses a statevector sampler so that the example can be
reproduced locally. An exact minimum eigensolver is included as a reference
for checking the result of this small instance.

\begin{lstlisting}[style=Pythonstyle]
import numpy as np

from qiskit.primitives import StatevectorSampler

from qiskit_finance.applications.optimization import (
    PortfolioOptimization
)

from qiskit_optimization.algorithms import (
    MinimumEigenOptimizer
)

from qiskit_optimization.minimum_eigensolvers import (
    QAOA,
    NumPyMinimumEigensolver
)

from qiskit_optimization.optimizers import COBYLA

# Illustrative financial inputs
expected_returns = np.array([
    0.10,
    0.20,
    0.15
])

covariances = np.array([
    [0.10, 0.02, 0.01],
    [0.02, 0.08, 0.03],
    [0.01, 0.03, 0.09]
])

risk_factor = 0.5
budget = 2

# Construct the portfolio problem
portfolio = PortfolioOptimization(
    expected_returns=expected_returns,
    covariances=covariances,
    risk_factor=risk_factor,
    budget=budget
)

qp = portfolio.to_quadratic_program()

# QAOA solver
sampler = StatevectorSampler(
    seed=123
)

qaoa = QAOA(
    sampler=sampler,
    optimizer=COBYLA(
        maxiter=200
    ),
    reps=2,
    initial_point=[
        0.5,
        0.5,
        0.5,
        0.5
    ]
)

qaoa_solver = MinimumEigenOptimizer(
    qaoa
)

qaoa_result = qaoa_solver.solve(
    qp
)

# Exact reference solution
exact_solver = MinimumEigenOptimizer(
    NumPyMinimumEigensolver()
)

exact_result = exact_solver.solve(
    qp
)

print("QAOA selection:")
print(qaoa_result.x)

print("Exact selection:")
print(exact_result.x)
\end{lstlisting}

The \texttt{PortfolioOptimization} class first constructs the constrained
quadratic problem from the expected returns, covariance matrix, risk
parameter, and budget. \texttt{MinimumEigenOptimizer} performs the
conversion required by the minimum eigensolver, including the treatment of
the constraint when the problem is mapped to QUBO form. The QAOA instance
uses two cost and mixer layers. The exact solver is practical only because
this example contains three binary variables and is included to provide a
known reference solution.

The structure of the two layer QAOA circuit is shown in
Fig.~\ref{fig:portfolio_qaoa}. The detailed decomposition of the cost
operation depends on the coefficients generated from the expected returns,
covariances, and budget penalty. It is therefore represented as a complete
cost unitary instead of assigning arbitrary \(R_z\) and \(ZZ\) gates to
individual financial terms.

\begin{figure}
\centering
\begin{quantikz}[column sep=0.55cm, row sep=0.45cm]
\lstick{$\ket{0}_{x_0}$} & \gate{H} & \gate[wires=3]{U_C(\gamma_1)} & \gate{R_x(2\beta_1)} & \gate[wires=3]{U_C(\gamma_2)} & \gate{R_x(2\beta_2)} & \meter{} \\
\lstick{$\ket{0}_{x_1}$} & \gate{H} & & \gate{R_x(2\beta_1)} & & \gate{R_x(2\beta_2)} & \meter{} \\
\lstick{$\ket{0}_{x_2}$} & \gate{H} & & \gate{R_x(2\beta_1)} & & \gate{R_x(2\beta_2)} & \meter{}
\end{quantikz}

\caption{Two layer QAOA circuit for a portfolio containing three binary
asset selection variables. The cost operations
\(U_C(\gamma_1)\) and \(U_C(\gamma_2)\) encode the Ising representation of
the portfolio objective and its budget penalty. The mixer operations are
implemented through \(R_x\) rotations controlled by
\(\beta_1\) and \(\beta_2\). Measurement produces candidate asset
selection bit strings.}
\label{fig:portfolio_qaoa}
\end{figure}

In this circuit, the \(H\) gates prepare a superposition of the eight
possible selection strings for three assets. The cost operation changes
their phases according to the portfolio objective. The mixer then
redistributes amplitude among the computational basis states. This sequence
is repeated because \texttt{reps=2} in the example. The parameters
\(\gamma_1\) and \(\gamma_2\) control the two cost operations, while
\(\beta_1\) and \(\beta_2\) control the corresponding mixers. These four
parameters are updated by constrained optimization by linear approximation (COBYLA) during the classical optimization loop.

Measurement returns a distribution over binary strings rather than a
guaranteed optimum from a single circuit execution. For example, a string
such as \(011\) represents the selection of two of the three assets under
the convention used by the optimization variables. QAOA attempts to place
greater probability on strings with low objective values, but its success
depends on the problem instance, circuit depth, optimizer, parameter
initialization, and execution noise. The exact eigensolver in the code
allows these effects to be separated from errors in the financial
formulation for this small example.

Buonaiuto \textit{et al.}~\cite{buonaiuto2023best} examined portfolio
optimization with variational quantum eigensolver (VQE) using \textit{Qiskit} and placed particular emphasis on the
choice of circuit and optimizer. Their constrained portfolio formulation
was converted to QUBO form and then to an Ising Hamiltonian. The authors
compared the \texttt{TwoLocal}, \texttt{RealAmplitudes}, and
\texttt{PauliTwoDesign} circuit structures together with several classical
optimizers. The experiments included ideal simulation, simulation with
noise, and execution on physical IBM quantum processors.

The study found that circuit structure and optimizer choice had a measurable
effect on convergence. COBYLA was among the more stable optimizers in the
experiments, while the behaviour of the different circuit structures
changed when noise was introduced. The authors reported that suitable
configurations on the tested physical devices produced solutions close to
the exact solution for their small portfolio instances. Noise generally
slowed convergence or increased the minimum energy obtained by the VQE.

These experiments provide useful evidence about parameter selection for
small VQE portfolio problems. They do not establish that VQE is currently
more efficient than classical portfolio optimization. The authors relate
the possibility of greater efficiency to future quantum processors with
sufficiently large usable registers. The present hardware experiments are
therefore most informative as studies of circuit configuration and noise
sensitivity.

Carrascal \textit{et al.}~\cite{carrascal2023backtesting} examined the
financial behaviour of quantum portfolio strategies using historical
IBEX35 data. Their comparison included VQE, a Conditional Value at Risk (CVaR)
variant of VQE, QAOA, and Grover Adaptive Search, together with three
classical strategies. The implementations were developed in Python using
\textit{Qiskit}, and the study performed 10,000 backtesting experiments under common
portfolio conditions.

The historical backtest used market data from 2016 to 2020. The strategies
were periodically applied to asset subsets so that portfolio selections
could be compared under the same experimental procedure. The authors report
that the quantum strategies matched or slightly exceeded the selected
classical strategies in parts of their backtesting analysis. These results
refer to portfolio performance under the specific historical dataset,
selection rules, and benchmark algorithms used in the experiment. They are
not measurements of quantum computational speedup.

The same study also examined VQE circuits on physical IBM processors with
27 and 127 qubits. A 28 qubit VQE circuit was mapped to part of the
127 qubit IBM Washington processor, and the authors reported 48.6 seconds
for 100,000 shots on the quantum system for that experiment. Queueing and
other service delays were analysed separately because wall clock job time
can be much larger than circuit execution time. These hardware experiments
show that relatively wide portfolio circuits can be executed, but the
backtesting results themselves do not demonstrate that the hardware
implementation is faster than a classical optimizer.

Carrascal \textit{et al.}~\cite{carrascal2024differential} studied a
different financial optimization problem involving cryptocurrency
arbitrage. The problem was formulated as a constrained binary optimization
model and converted to an Ising representation for VQE using \textit{Qiskit}. The
main algorithmic contribution was the use of Differential Evolution as the
classical optimizer for the variational circuit. The authors compared this
population based optimizer with alternatives such as COBYLA in a landscape
containing several local minima.

For the three currency example, both simulation and physical IBM hardware
were considered. The authors report convergence after 417 optimization
steps over approximately 12 hours on the
\texttt{ibm\_geneva} processor. They also tested larger cryptocurrency
instances on other IBM processors, including systems with up to 127 physical
qubits. The larger real hardware experiments were considerably more
expensive. For the five currency case, hardware tests were stopped before
convergence because completing the number of generations suggested by
simulation would require substantially longer execution.

The study therefore provides a direct \textit{Qiskit} hardware experiment for a
financial optimization problem and shows how a population based classical
optimizer can be incorporated into a variational workflow. The reported
twelve hour convergence time for the three currency example also shows that
successful optimization on current hardware does not imply a practical
runtime advantage. The authors present the experiments as evidence of
feasibility and discuss quantum advantage as a possible future outcome as
hardware and parallel execution improve.

Yalovetzky \textit{et al.}~\cite{yalovetzky2021nisq} considered portfolio
optimization through a different mathematical formulation. Their noisy intermediate scale quantum (NISQ) Harrow Hassidim Lloyd (HHL)
method converts a mean variance portfolio problem into a quantum linear
system. If the resulting system is written as
\(A\mathbf{z}=\mathbf{b}\), then \(A\) contains the coefficients of the
linear system, \(\mathbf{b}\) contains the corresponding constraints and
target terms, and \(\mathbf{z}\) represents the solution from which the
portfolio allocation is obtained.

The authors developed the circuits in \textit{Qiskit} and used \textit{Qiskit} Aer statevector and quantum assembly language (QASM) simulation during validation. Their modified phase
estimation procedure uses mid circuit measurement, qubit reset, and quantum
conditional logic so that a single ancillary qubit can be reused during
eigenvalue estimation. For execution on the trapped ion Honeywell System
Model H1, the circuits constructed in \textit{Qiskit} were translated to the
device's native operations using \texttt{pytket}. The hardware experiment
therefore combines \textit{Qiskit} circuit construction with a different hardware
software stack.

The physical experiment considered a portfolio containing two S\&P 500
assets, which produced a \(4\times4\) linear system after the constraints
were incorporated. In one comparison, the inner product obtained using the
\textit{Qiskit} statevector simulator was approximately 0.83 for the NISQ HHL
construction and 0.59 for the earlier uniformly controlled rotation
method. On the H1 hardware, the corresponding values were approximately
0.46 and 0.42. The reduction from simulation to hardware reflects the
depth of the circuits and the effect of device noise.

The same work also examined portfolios containing six and fourteen assets
using \textit{Qiskit} statevector simulation. These larger examples were not
end to end hardware demonstrations. The study is therefore useful for
showing how \textit{Qiskit} supported the design and simulation of a portfolio
linear solver while hardware execution required additional tooling and
remained limited to a small portfolio instance.

The portfolio studies reviewed here use \textit{Qiskit} in several distinct ways.
QAOA and VQE treat asset selection as an optimization problem represented
by a QUBO or Ising Hamiltonian. Historical backtesting evaluates the
financial behaviour of the resulting portfolio decisions, while the
NISQ HHL work approaches mean variance optimization through a linear
system formulation. Physical processor experiments have been reported for
all of these broader approaches, but the demonstrated instances remain
small or require substantial execution time. Current results establish that
\textit{Qiskit} can support complete experimental workflows for portfolio
optimization, while a computational advantage over mature classical
portfolio solvers has not yet been demonstrated by the studies considered
here.

\subsubsection{Risk Management}

Portfolio optimization determines which assets should be selected, while
risk management considers the losses that may occur after those positions
have been taken. \textit{Qiskit} based studies in this area have examined market
risk, counterparty credit risk, portfolio loss distributions, and financial
fraud detection. A recurring method is quantum amplitude estimation, which
can be used to estimate probabilities and expectations encoded in quantum
states.

VaR and CVaR summarize different parts of a
loss distribution. Let \(L\) denote a random variable representing financial
loss and let \(c\in(0,1)\) denote a confidence level. These measures can be
written as

\begin{equation}
\begin{aligned}
\mathrm{VaR}_{c}(L)
&=
\inf
\left\{
\ell:
\Pr(L\leq\ell)\geq c
\right\}, \\
\mathrm{CVaR}_{c}(L)
&=
\mathbb{E}
\left[
L
\mid
L\geq\mathrm{VaR}_{c}(L)
\right].
\end{aligned}
\end{equation}

Here, \(\mathrm{VaR}_{c}(L)\) is the loss threshold associated with
confidence level \(c\), while \(\mathrm{CVaR}_{c}(L)\) is the expected loss
in the tail beyond that threshold~\cite{best2000implementing,chow2015risk}.
For discrete loss distributions, the treatment of probability mass exactly
at the threshold may require an adjusted tail average, but the expressions
above give the risk measures used throughout this discussion.

Quantum amplitude estimation can be applied after a loss distribution has
been encoded into a quantum state. If a circuit prepares the possible loss
levels with their corresponding probabilities and an objective qubit marks
the outcomes above a chosen threshold, amplitude estimation returns an
estimate of the probability associated with that marked region. Repeating
the procedure while changing the threshold allows a VaR level to be located.
A related amplitude encoding can then be used to estimate the expected loss
in the tail for CVaR~\cite{QF}.

The following example illustrates the probability estimation step using a
small discrete loss distribution. Four possible loss levels are represented
using two qubits. The probabilities are illustrative and sum to one. The
objective qubit is set to one when the encoded loss is at least two. Since
the computational basis states \(\lvert10\rangle\) and
\(\lvert11\rangle\) correspond to loss levels two and three, the most
significant loss qubit can be used directly to mark the required tail.

\begin{lstlisting}[style=Pythonstyle]
import numpy as np

from qiskit import QuantumCircuit
from qiskit.circuit.library import StatePreparation
from qiskit.primitives import StatevectorSampler

from qiskit_algorithms import (
    EstimationProblem,
    IterativeAmplitudeEstimation
)

# Discrete loss distribution
losses = np.array([
    0.0,
    1.0,
    2.0,
    3.0
])

probabilities = np.array([
    0.55,
    0.25,
    0.15,
    0.05
])

threshold = 2.0

# Amplitudes encode the loss probabilities
amplitudes = np.sqrt(probabilities)

# Two loss qubits and one objective qubit
state_preparation = QuantumCircuit(3)

state_preparation.append(
    StatePreparation(amplitudes),
    [0, 1]
)

# q1 is one for states |10> and |11>
state_preparation.cx(1, 2)

problem = EstimationProblem(
    state_preparation=state_preparation,
    objective_qubits=[2]
)

sampler = StatevectorSampler(
    default_shots=4096,
    seed=123
)

iae = IterativeAmplitudeEstimation(
    epsilon_target=0.01,
    alpha=0.05,
    sampler=sampler
)

result = iae.estimate(problem)

exact_tail_probability = probabilities[
    losses >= threshold
].sum()

print(
    "Estimated tail probability:",
    result.estimation
)

print(
    "Exact tail probability:",
    exact_tail_probability
)
\end{lstlisting}

The state preparation operation assigns amplitude
\(\sqrt{p_j}\) to each computational basis state, where \(p_j\) is the
probability associated with loss level \(j\). In this example, the four
probabilities correspond to losses zero, one, two, and three. The final
controlled operation marks the two states belonging to the selected tail.
The exact probability of this tail is \(0.15+0.05=0.20\). Iterative
amplitude estimation estimates this probability from repeated applications
of the state preparation and associated Grover operations.

The logical circuit used for the state preparation and tail marking step is
shown in Fig.~\ref{fig:risk_iae}. The block
\(\mathcal{A}_{L}\) prepares the discrete loss distribution on the two loss
qubits. The objective qubit \(o\) records whether the encoded loss belongs
to the selected tail.

\begin{figure}
\centering
\begin{quantikz}[column sep=0.8cm, row sep=0.5cm]
\lstick{$\ket{0}_{\ell_0}$} & \gate[wires=2]{\mathcal{A}_{L}} & \qw & \qw \\
\lstick{$\ket{0}_{\ell_1}$} & & \ctrl{1} & \qw \\
\lstick{$\ket{0}_{o}$} & \qw & \targ{} & \meter{}
\end{quantikz}

\caption{Illustrative loss probability circuit used as the state preparation
step for amplitude estimation. The operation
\(\mathcal{A}_{L}\) prepares the discrete loss distribution on
\(\ell_0\) and \(\ell_1\). For the threshold used in the example,
\(\ell_1=1\) identifies the loss states \(\lvert10\rangle\) and
\(\lvert11\rangle\), so the objective qubit is set to one for those
outcomes. Iterative amplitude estimation uses this marked probability to
estimate the probability of a loss at or above the selected threshold.}
\label{fig:risk_iae}
\end{figure}

The circuit in Fig.~\ref{fig:risk_iae} represents only the small state
preparation example used to explain the method. Practical risk models require
larger uncertainty registers, correlated risk factors, arithmetic for
aggregating losses, and circuits that identify the required quantile.
\textit{Qiskit} Finance provides a credit risk example in which a Gaussian
conditional independence model is used to represent correlated defaults,
losses are aggregated, and iterative amplitude estimation is applied to the
resulting distribution~\cite{QF}. The theoretical improvement associated
with amplitude estimation concerns the number of oracle queries required
for a specified estimation precision. It should not be interpreted as a
guaranteed reduction in total execution time on present quantum hardware.

Wilkens and Moorhouse~\cite{wilkens2023quantum} examined this distinction
in the context of financial institutions. Their \textit{Qiskit} implementation
considered both Value at Risk for market risk and Potential Future Exposure
for counterparty credit risk. For market risk, future changes in financial
risk factors were represented by quantum probability distributions and
combined to obtain a portfolio profit and loss distribution. The resulting
quantile was then estimated using amplitude estimation. Their counterparty
credit risk model extended the procedure across multiple future time
points, where exposures must be calculated repeatedly as market conditions
change.

The experiments were performed using quantum simulation rather than a
physical processor. The authors considered realistic requirements such as
non Gaussian marginal distributions, dependence between risk factors, and
multiple future horizons. Their analysis showed that these requirements
cause the number of required qubits and circuit operations to grow rapidly.
For the counterparty credit risk example, they also introduced a simplified
noise model in which one and two qubit operations and measurements were
subject to errors. With three qubits used for each risk factor, the resulting
errors in some Potential Future Exposure quantiles exceeded 25\%.

The authors therefore reached a cautious conclusion. Small risk measurement
circuits can be constructed in \textit{Qiskit}, but the number of risk factors in a
real financial institution is much larger than the examples that can be
represented with current hardware. They estimate that practical systems
would require substantially greater qubit capacity, together with improved
noise control. Their analysis does not present quantum risk measurement as
a current replacement for established classical systems and notes that the
business case remains limited while classical implementations already
provide sufficient accuracy and execution speed for many institutional risk
calculations~\cite{wilkens2023quantum}.

Dri \textit{et al.}~\cite{dri2023general} examined credit risk in more
detail by extending the commonly used single factor default model to include
multiple systemic risk factors. Each counterparty has a probability of
default that depends on the realization of these factors. The authors
considered two quantum circuit constructions. One assigns a separate
quantum register to each systemic factor and applies multiple controlled
rotations to the counterparty qubits. The other encodes the factors using a
multivariate distribution and combines their effects before applying the
default rotations.

Their noiseless \textit{Qiskit} experiment contained two assets and two systemic risk
factors. Two qubits were used to represent each Gaussian risk factor, and
the complete circuit required nine qubits. VaR was estimated at a confidence
level of 95\% using iterative amplitude estimation with a target estimation
precision of \(0.002\) and a 99\% target confidence interval. The authors
report that approximately 50,000 quantum samples were required on average
under this configuration.

The same work also investigated noisy simulation and physical IBM quantum
processors. The tested configurations contained between two and four assets
and between one and three systemic risk factors, requiring between seven and
thirteen qubits. The hardware experiments included the IBM Perth and IBM
Lagos processors. Noise had a substantial effect as circuit size increased,
particularly because the representation of additional factors requires
further controlled rotations and therefore greater circuit depth.

The contribution of this study is the construction of a more flexible credit
risk model that can represent several sources of systemic dependence rather
than the demonstration of a production scale risk engine. The largest
experiments still contain only a few counterparties, whereas institutional
credit portfolios may contain many thousands. The \textit{Qiskit} experiments
therefore validate the circuit construction and expose its resource
requirements, but they do not demonstrate a practical computational
advantage for large credit portfolios.

de Pedro \textit{et al.}~\cite{de2023var} considered market VaR from a
different direction. Instead of encoding every asset in a separate quantum
register, they used random variable algebra and a Taylor approximation to
combine the contributions of the assets before preparing the final
distribution. This allows the number of assets in the financial portfolio
to be much larger than the number of qubits used to represent the resulting
risk distribution.

Their \textit{Qiskit} implementation first restricts the prepared distribution to a
window containing the required tail. For a four qubit example, the selected
window is divided into sixteen discrete values. A state preparation circuit
then assigns amplitudes according to the approximated probability
distribution. The measured computational basis state is converted back to
a financial loss using the interval represented by the register.

The authors tested the circuit using \textit{Qiskit} statevector simulation and later
on the IBM Lagos processor. For one reported portfolio containing
138 assets, the four qubit simulator produced a VaR estimate of
United States Dollar (USD) \(-9111.50\), compared with a parametric reference value of
USD \(-9514.18\). The physical hardware distribution was considerably
distorted by noise, showing that a state which is useful under ideal
simulation may not retain the same distribution after execution on a
current processor.

To address this problem, the authors introduced a neural network based
correction procedure that adjusts the circuit parameters using the
difference between the desired and measured distributions. Across their
experiments, portfolios containing up to 139 assets were considered. The
authors report empirical VaR errors below 1\% relative to their parametric
reference after applying the correction procedure in the tested cases.
This result does not mean that 139 assets were individually represented by
139 qubits. The portfolio is first reduced analytically to an aggregated
risk distribution, which is then represented using a small quantum
register. The result therefore demonstrates a compact approach to a
particular VaR calculation rather than a direct quantum representation of
all individual positions.

Risk management also includes identifying transactions that may indicate
fraud. Grossi \textit{et al.}~\cite{grossi2022mixed} developed a mixed
quantum and classical fraud detection workflow using real card payment data,
IBM Safer Payments, IBM Quantum systems, and the \textit{Qiskit} software stack. Their
quantum component used a Quantum SVM (QSVM) together with a
quantum feature selection procedure. Classical Random Forest and XGBoost
models were included as comparison methods.

The authors found that the behaviour of the quantum classifier depended
strongly on the features selected for the experiment. The available quantum
hardware also required the original payment dataset to be reduced
substantially before the quantum model could be evaluated. Rather than
reporting the QSVM as a general replacement for the classical classifiers,
the study examined whether the quantum feature representation could provide
additional information. A mixed model combining quantum and classical
outputs improved selected fraud detection measures for the reduced dataset
used in the experiment.

This distinction is important because fraud detection systems used by
financial institutions operate on substantially larger and continuously
changing transaction streams. The study demonstrates an end to end \textit{Qiskit}
workflow on real financial payment data, but its results do not establish
that the quantum classifier can currently scale to production transaction
volumes or provide lower inference latency than established classical
systems.

The risk studies discussed above cover three different uses of \textit{Qiskit}.
Amplitude estimation is used to estimate quantities derived from market and
credit loss distributions. Quantum state preparation and noise correction
are used to study VaR on physical hardware, while quantum kernels are used
for classification in fraud detection. The experiments range from ideal
simulation to physical quantum processors and therefore provide different
levels of evidence. They show that the required financial quantities can be
represented by quantum circuits and that selected small instances can be
executed on current hardware. They also identify significant limitations in
distribution loading, circuit depth, noise, available qubits, and dataset
size. The studies considered here consequently do not demonstrate that
\textit{Qiskit} based risk calculations currently outperform established classical
risk systems at institutional scale.

\section{Conclusion} \label{sec:conclusion}

In this paper, we present a structured and detailed review of \textit{Qiskit}, focusing on its evolution, contemporary architecture, and use across diverse application domains. We trace the framework from its earlier element architecture, including Terra, Aer, Ignis, and Aqua, to the current \textit{Qiskit} 2.x environment built around the SDK, separately maintained ecosystem packages, primitives, simulation tools, and IBM Quantum execution services. These changes are important when interpreting published applications because several interfaces and programming patterns used in earlier studies are no longer compatible with current \textit{Qiskit} versions. The application literature is examined across cryptography and cybersecurity, image and signal processing, climate science and energy applications, and finance and risk analysis, where \textit{Qiskit} has been used to construct quantum circuits, develop hybrid classical quantum workflows, perform ideal and noisy simulation, and execute selected experiments on physical quantum processors. The reviewed studies include QKD and random number generation, medical imaging and telecommunications, environmental modelling and energy optimization, and portfolio optimization and financial risk estimation. Across these areas, the evidence also reveals important limitations, as many reported experiments remain small in scale, rely substantially on simulation or reduced representations of classical data, and are affected by circuit depth, hardware noise, execution cost, and limited quantum resources. Physical hardware demonstrations establish that selected workflows can be executed on current processors, but they do not by themselves demonstrate a computational advantage over established classical methods. Taken together, the reviewed literature shows that \textit{Qiskit} provides a broad experimental environment for developing and evaluating quantum algorithms, while further progress in hardware capability, software stability, and comparative evaluation is required before many of these applications can be demonstrated at practically meaningful scales.

\section{List of Abbreviations} 

\begin{itemize}[left=0pt, itemsep=2pt, parsep=0pt, topsep=0pt, partopsep=0pt,label={}]

\item \makebox[8em][l]{\textbf{API}} Application Programming Interface
\item \makebox[8em][l]{\textbf{AUC}} Area Under the Curve
\item \makebox[8em][l]{\textbf{BB84}} Bennett Brassard 1984
\item \makebox[8em][l]{\textbf{BER}} Bit Error Rate
\item \makebox[8em][l]{\textbf{CHSH}} Clauser Horne Shimony Holt
\item \makebox[8em][l]{\textbf{CNOT}} Controlled NOT
\item \makebox[8em][l]{\textbf{COBYLA}} Constrained Optimization by Linear Approximation
\item \makebox[8em][l]{\textbf{CVaR}} Conditional Value at Risk
\item \makebox[8em][l]{\textbf{FRQI}} Flexible Representation of Quantum Images
\item \makebox[8em][l]{\textbf{\(H\)}} Hadamard
\item \makebox[8em][l]{\textbf{HHL}} Harrow Hassidim Lloyd
\item \makebox[8em][l]{\textbf{IBM}} International Business Machines
\item \makebox[8em][l]{\textbf{IQFT}} Inverse Quantum Fourier Transform
\item \makebox[8em][l]{\textbf{MD5}} Message Digest Algorithm 5
\item \makebox[8em][l]{\textbf{MIMO}} Multiple Input Multiple Output
\item \makebox[8em][l]{\textbf{NISQ}} Noisy Intermediate Scale Quantum
\item \makebox[8em][l]{\textbf{NIST}} National Institute of Standards and Technology
\item \makebox[8em][l]{\textbf{NOMA}} Non Orthogonal Multiple Access
\item \makebox[8em][l]{\textbf{NPCC}} Northeast Power Coordinating Council
\item \makebox[8em][l]{\textbf{OFDM}} Orthogonal Frequency Division Multiplexing
\item \makebox[8em][l]{\textbf{PQC}} Parameterized Quantum Circuit
\item \makebox[8em][l]{\textbf{PUB}} Primitive Unified Bloc
\item \makebox[8em][l]{\textbf{QAOA}} Quantum Approximate Optimization Algorithm
\item \makebox[8em][l]{\textbf{QASM}} Quantum Assembly Language
\item \makebox[8em][l]{\textbf{QBER}} Quantum Bit Error Rate
\item \makebox[8em][l]{\textbf{QC}} Quantum Computing
\item \makebox[8em][l]{\textbf{QCNN}} Quantum Convolutional Neural Network
\item \makebox[8em][l]{\textbf{QFT}} Quantum Fourier Transform
\item \makebox[8em][l]{\textbf{QKD}} Quantum Key Distribution
\item \makebox[8em][l]{\textbf{QML}} Quantum Machine Learning
\item \makebox[8em][l]{\textbf{QRNG}} Quantum Random Number Generation / Generator
\item \makebox[8em][l]{\textbf{QSVC}} Quantum Support Vector Classifier
\item \makebox[8em][l]{\textbf{QSVM}} Quantum Support Vector Machine
\item \makebox[8em][l]{\textbf{QUBO}} Quadratic Unconstrained Binary Optimization
\item \makebox[8em][l]{\textbf{ROC}} Receiver Operating Characteristic
\item \makebox[8em][l]{\textbf{SDK}} Software Development Kit
\item \makebox[8em][l]{\textbf{SHA}} Secure Hash Algorithm
\item \makebox[8em][l]{\textbf{SVM}} Support Vector Machine
\item \makebox[8em][l]{\textbf{USD}} United States Dollar
\item \makebox[8em][l]{\textbf{VaR}} Value at Risk
\item \makebox[8em][l]{\textbf{VQC}} Variational Quantum Classifier
\item \makebox[8em][l]{\textbf{VQE}} Variational Quantum Eigensolver
\item \makebox[8em][l]{\textbf{VQR}} Variational Quantum Regressor

\end{itemize}

\bibliographystyle{unsrt}
\bibliography{references}

@book{nielsen2010quantum,
  author = {Nielsen, Michael A. and Chuang, Isaac L.},
  title = {Quantum Computation and Quantum Information: 10th Anniversary Edition},
  publisher = {Cambridge University Press},
  year = {2010},
  doi = {10.1017/CBO9780511976667},
  url = {https://doi.org/10.1017/CBO9780511976667}
}

@article{shor1999polynomial,
  author = {Shor, Peter W.},
  title = {Polynomial-Time Algorithms for Prime Factorization and Discrete Logarithms on a Quantum Computer},
  journal = {SIAM Review},
  volume = {41},
  number = {2},
  pages = {303--332},
  year = {1999},
  doi = {10.1137/S0036144598347011},
  url = {https://doi.org/10.1137/S0036144598347011}
}

@inproceedings{shor1994algorithms,
  author = {Shor, Peter W.},
  title = {Algorithms for Quantum Computation: Discrete Logarithms and Factoring},
  booktitle = {Proceedings of the 35th Annual Symposium on Foundations of Computer Science},
  pages = {124--134},
  year = {1994},
  publisher = {IEEE},
  doi = {10.1109/SFCS.1994.365700},
  url = {https://doi.org/10.1109/SFCS.1994.365700}
}

@inproceedings{grover1996fast,
  author = {Grover, Lov K.},
  title = {A Fast Quantum Mechanical Algorithm for Database Search},
  booktitle = {Proceedings of the Twenty-Eighth Annual ACM Symposium on Theory of Computing},
  pages = {212--219},
  year = {1996},
  publisher = {ACM},
  doi = {10.1145/237814.237866},
  url = {https://doi.org/10.1145/237814.237866}
}

@misc{javadi2024quantum,
  author = {Javadi-Abhari, Ali and Treinish, Matthew and Krsulich, Kevin and Wood, Christopher J. and Lishman, Jake and Gacon, Julien and Martiel, Simon and Nation, Paul D. and Bishop, Lev S. and Cross, Andrew W. and Johnson, Blake R. and Gambetta, Jay M.},
  title = {Quantum Computing with {Qiskit}},
  year = {2024},
  eprint = {2405.08810},
  archivePrefix = {arXiv},
  primaryClass = {quant-ph},
  doi = {10.48550/arXiv.2405.08810},
  url = {https://arxiv.org/abs/2405.08810}
}

@article{fingerhuth2018open,
  author = {Fingerhuth, Mark and Babej, Tom{\'a}{\v{s}} and Wittek, Peter},
  title = {Open Source Software in Quantum Computing},
  journal = {PLOS ONE},
  volume = {13},
  number = {12},
  pages = {e0208561},
  year = {2018},
  doi = {10.1371/journal.pone.0208561},
  url = {https://doi.org/10.1371/journal.pone.0208561}
}

@misc{sahin2025qiskit,
  author = {Sahin, M. Emre and Altamura, Edoardo and Wallis, Oscar and Wood, Stephen P. and Dekusar, Anton and Millar, Declan A. and Imamichi, Takashi and Matsuo, Atsushi and Mensa, Stefano},
  title = {{Qiskit} Machine Learning: An Open-Source Library for Quantum Machine Learning Tasks at Scale on Quantum Hardware and Classical Simulators},
  year = {2025},
  eprint = {2505.17756},
  archivePrefix = {arXiv},
  primaryClass = {quant-ph},
  doi = {10.48550/arXiv.2505.17756},
  url = {https://arxiv.org/abs/2505.17756}
}

@inproceedings{stirbu2023quantum,
  author = {Stirbu, Vlad and Mikkonen, Tommi},
  title = {Quantum Software Ecosystem: Stakeholders, Interactions and Challenges},
  booktitle = {Software Business. ICSOB 2023},
  series = {Lecture Notes in Business Information Processing},
  volume = {500},
  pages = {471--477},
  year = {2024},
  publisher = {Springer},
  doi = {10.1007/978-3-031-53227-6_33},
  url = {https://doi.org/10.1007/978-3-031-53227-6_33}
}

@article{seskir2022quantum,
  author = {Seskir, Zeki C. and Migda{\l}, Piotr and Weidner, Carrie and Anupam, Aditya and Case, Nicky and Davis, Noah and Decaroli, Chiara and Ercan, {\.I}lke and Foti, Caterina and Gora, Pawe{\l} and others},
  title = {Quantum Games and Interactive Tools for Quantum Technologies Outreach and Education},
  journal = {Optical Engineering},
  volume = {61},
  number = {8},
  pages = {081809},
  year = {2022},
  doi = {10.1117/1.OE.61.8.081809},
  url = {https://doi.org/10.1117/1.OE.61.8.081809}
}

@article{alexander2020qiskit,
  author = {Alexander, Thomas and Kanazawa, Naoki and Egger, Daniel J. and Capelluto, Lauren and Wood, Christopher J. and Javadi-Abhari, Ali and McKay, David C.},
  title = {{Qiskit} Pulse: Programming Quantum Computers Through the Cloud with Pulses},
  journal = {Quantum Science and Technology},
  volume = {5},
  number = {4},
  pages = {044006},
  year = {2020},
  doi = {10.1088/2058-9565/aba404},
  url = {https://doi.org/10.1088/2058-9565/aba404}
}

@misc{ibmhardware,
  author = {{IBM Quantum}},
  title = {{IBM Quantum Computing: Hardware and Roadmap}},
  year = {2026},
  note = {Accessed 28 August 2026},
  url = {https://www.ibm.com/quantum/hardware}
}

@misc{ibmroadmap,
  author = {{IBM}},
  title = {{Quantum Roadmap: Strategic Milestones}},
  year = {2026},
  note = {IBM Technology Atlas. Accessed 28 August 2026},
  url = {https://www.ibm.com/roadmaps/quantum/}
}

@article{shammah2023open,
  author = {Shammah, Nathan and Saha Roy, Anurag and Almudever, Carmen G. and Bourdeauducq, S{\'e}bastien and Butko, Anastasiia and Cancelo, Gustavo and Clark, Susan M. and Heinsoo, Johannes and Henriet, Lo{\"\i}c and Huang, Gang and Jurczak, Christophe and Kotilahti, Janne and Landra, Alessandro and LaRose, Ryan and Mari, Andrea and Nowrouzi, Kasra and Ockeloen-Korppi, Caspar and Prawiroatmodjo, Guen and Siddiqi, Irfan and Zeng, William J.},
  title = {Open Hardware Solutions in Quantum Technology},
  journal = {APL Quantum},
  volume = {1},
  number = {1},
  pages = {011501},
  year = {2024},
  doi = {10.1063/5.0180987},
  url = {https://doi.org/10.1063/5.0180987}
}

@misc{qiskit044notes,
  author = {{Qiskit Development Team}},
  title = {{Qiskit 0.44 Release Notes}},
  year = {2023},
  note = {IBM Quantum Documentation. Accessed 28 August 2026},
  url = {https://quantum.cloud.ibm.com/docs/en/api/qiskit/release-notes/0.44}
}

@misc{qiskitaqua,
  author = {{Qiskit Community}},
  title = {{Qiskit Aqua: Quantum Algorithms and Applications}},
  year = {2021},
  note = {Deprecated repository. Accessed 28 August 2026},
  url = {https://github.com/qiskit-community/qiskit-aqua}
}

@misc{qiskit033notes,
  author = {{Qiskit Development Team}},
  title = {{Qiskit 0.33 Release Notes}},
  year = {2021},
  note = {IBM Quantum Documentation. Accessed 28 August 2026},
  url = {https://quantum.cloud.ibm.com/docs/en/api/qiskit/release-notes/0.33}
}

@misc{qiskitignis,
  author = {{Qiskit Community}},
  title = {{Qiskit Ignis}},
  year = {2021},
  note = {Deprecated repository. Accessed 28 August 2026},
  url = {https://github.com/qiskit-community/qiskit-ignis}
}

@misc{qiskit040notes,
  author = {{Qiskit Development Team}},
  title = {{Qiskit 0.40 Release Notes}},
  year = {2023},
  note = {IBM Quantum Documentation. Accessed 28 August 2026},
  url = {https://quantum.cloud.ibm.com/docs/en/api/qiskit/release-notes/0.40}
}

@misc{qiskit045notes,
  author = {{Qiskit Development Team}},
  title = {{Qiskit 0.45 Release Notes}},
  year = {2023},
  note = {IBM Quantum Documentation. Accessed 28 August 2026},
  url = {https://quantum.cloud.ibm.com/docs/en/api/qiskit/release-notes/0.45}
}

@misc{qiskitv2primitives,
  author = {{IBM Quantum}},
  title = {{Migrate to the Qiskit Runtime V2 Primitives}},
  year = {2024},
  note = {IBM Quantum Documentation. Accessed 28 August 2026},
  url = {https://quantum.cloud.ibm.com/docs/en/migration-guides/v2-primitives}
}

@misc{qiskit1launch,
  author = {Johnson, Blake and Treinish, Matthew and Bello, Luciano and Lishman, Jake and Krsulich, Kevin},
  title = {{Qiskit 1.0 Coming in February 2024}},
  year = {2023},
  month = dec,
  note = {IBM Quantum Blog. Accessed 28 August 2026},
  url = {https://www.ibm.com/quantum/blog/qiskit-1-launch}
}

@misc{qiskit046notes,
  author = {{Qiskit Development Team}},
  title = {{Qiskit 0.46 Release Notes}},
  year = {2024},
  note = {IBM Quantum Documentation. Accessed 28 August 2026},
  url = {https://quantum.cloud.ibm.com/docs/en/api/qiskit/release-notes/0.46}
}

@misc{qiskit1features,
  author = {{IBM Quantum}},
  title = {{Qiskit v1.0 Feature Changes}},
  year = {2024},
  note = {IBM Quantum Documentation. Accessed 28 August 2026},
  url = {https://quantum.cloud.ibm.com/docs/en/guides/qiskit-1.0-features}
}

@misc{qiskitroadmap,
  author = {{Qiskit Development Team}},
  title = {{Qiskit Roadmap and Version Support}},
  year = {2026},
  note = {GitHub Wiki. Accessed 28 August 2026},
  url = {https://github.com/Qiskit/qiskit/wiki/Roadmap}
}

@misc{qiskit23blog,
  author = {{Qiskit Team}},
  title = {{Release News: Qiskit v2.3 Is Here!}},
  year = {2026},
  month = jan,
  note = {IBM Quantum Blog. Accessed 28 August 2026},
  url = {https://www.ibm.com/quantum/blog/qiskit-2-3-release-summary}
}

@misc{qiskit24blog,
  author = {{Qiskit Team}},
  title = {{Release News: Qiskit v2.4 Is Here!}},
  year = {2026},
  month = apr,
  note = {IBM Quantum Blog. Accessed 28 August 2026},
  url = {https://www.ibm.com/quantum/blog/qiskit-2-4-release-summary}
}

@misc{qiskit25blog,
  author = {{Qiskit Team}},
  title = {{Release News: Qiskit v2.5 Is Here!}},
  year = {2026},
  month = jul,
  note = {IBM Quantum Blog. Accessed 28 August 2026},
  url = {https://www.ibm.com/quantum/blog/qiskit-2-5-release-summary}
}

@misc{qiskitfunctions,
  author = {{IBM Quantum}},
  title = {{Introduction to Qiskit Functions}},
  year = {2026},
  note = {IBM Quantum Documentation. Preview feature. Accessed 28 August 2026},
  url = {https://quantum.cloud.ibm.com/docs/en/guides/functions}
}

@article{renner2008security,
  author = {Renner, Renato},
  title = {Security of Quantum Key Distribution},
  journal = {International Journal of Quantum Information},
  volume = {6},
  number = {1},
  pages = {1--127},
  year = {2008},
  doi = {10.1142/S0219749908003256},
  url = {https://doi.org/10.1142/S0219749908003256}
}

@article{wootters1982single,
  author = {Wootters, William K. and Zurek, Wojciech H.},
  title = {A Single Quantum Cannot Be Cloned},
  journal = {Nature},
  volume = {299},
  number = {5886},
  pages = {802--803},
  year = {1982},
  doi = {10.1038/299802a0},
  url = {https://doi.org/10.1038/299802a0}
}

@article{shor2000simple,
  author = {Shor, Peter W. and Preskill, John},
  title = {Simple Proof of Security of the {BB84} Quantum Key Distribution Protocol},
  journal = {Physical Review Letters},
  volume = {85},
  number = {2},
  pages = {441--444},
  year = {2000},
  doi = {10.1103/PhysRevLett.85.441},
  url = {https://doi.org/10.1103/PhysRevLett.85.441}
}

@article{mohanty2024experimentally,
  author = {Mohanty, Tapaswini and Srivastava, Vikas and Debnath, Sumit Kumar and Roy, Debasish and Sakurai, Kouichi and Mukhopadhyay, Sourav},
  title = {An Experimentally Validated Feasible Quantum Protocol for Identity-Based Signature},
  journal = {S{\=a}dhan{\=a}},
  volume = {50},
  number = {1},
  pages = {23},
  year = {2025},
  doi = {10.1007/s12046-025-02676-3},
  url = {https://doi.org/10.1007/s12046-025-02676-3}
}

@article{faleiro2021device,
  author = {Faleiro, Ricardo and Goul{\~a}o, Manuel},
  title = {Device-Independent Quantum Authorization Based on the Clauser-Horne-Shimony-Holt Game},
  journal = {Physical Review A},
  volume = {103},
  number = {2},
  pages = {022430},
  year = {2021},
  doi = {10.1103/PhysRevA.103.022430},
  url = {https://doi.org/10.1103/PhysRevA.103.022430}
}

@misc{das2023design,
  author = {Das, Prodipto and Biswas, Sumit and Kanoo, Sandip and Podder, Veeradittya},
  title = {Design and Comparative Analysis of Quantum Hashing Algorithms Using {Qiskit}},
  year = {2023},
  howpublished = {TechRxiv preprint},
  doi = {10.36227/techrxiv.24037440.v1},
  url = {https://doi.org/10.36227/techrxiv.24037440.v1},
  note = {Preprint, posted 28 August 2023}
}

@article{kuang2022quantum,
  author = {Perepechaenko, Maria and Kuang, Randy},
  title = {Quantum Encryption and Decryption in {IBMQ} Systems Using Quantum Permutation Pad},
  journal = {Journal of Communications},
  volume = {17},
  number = {12},
  pages = {972--978},
  year = {2022},
  doi = {10.12720/jcm.17.12.972-978},
  url = {https://doi.org/10.12720/jcm.17.12.972-978}
}

@article{herrero2017quantum,
  author = {Herrero-Collantes, Miguel and Garcia-Escartin, Juan Carlos},
  title = {Quantum Random Number Generators},
  journal = {Reviews of Modern Physics},
  volume = {89},
  number = {1},
  pages = {015004},
  year = {2017},
  doi = {10.1103/RevModPhys.89.015004},
  url = {https://doi.org/10.1103/RevModPhys.89.015004}
}

@inproceedings{germain2022qubit,
  author = {Germain, Julie and Dantu, Ram and Thompson, Mark A.},
  title = {Qubit Reset and Refresh: A Gamechanger for Random Number Generation},
  booktitle = {Proceedings of the Twelfth ACM Conference on Data and Application Security and Privacy},
  pages = {367--369},
  year = {2022},
  publisher = {ACM},
  doi = {10.1145/3508398.3519364},
  url = {https://doi.org/10.1145/3508398.3519364}
}

@article{kumar2022quantum,
  author = {Kumar, Vaishnavi and Rayappan, John Bosco Balaguru and Amirtharajan, Rengarajan and Praveenkumar, Padmapriya},
  title = {Quantum True Random Number Generation on {IBM}'s Cloud Platform},
  journal = {Journal of King Saud University - Computer and Information Sciences},
  volume = {34},
  number = {8},
  pages = {6453--6465},
  year = {2022},
  doi = {10.1016/j.jksuci.2022.01.015},
  url = {https://doi.org/10.1016/j.jksuci.2022.01.015}
}

@article{li2021quantum,
  author = {Li, Yuanhao and Fei, Yangyang and Wang, Weilong and Meng, Xiangdong and Wang, Hong and Duan, Qianheng and Ma, Zhi},
  title = {Quantum Random Number Generator Using a Cloud Superconducting Quantum Computer Based on Source-Independent Protocol},
  journal = {Scientific Reports},
  volume = {11},
  number = {1},
  pages = {23873},
  year = {2021},
  doi = {10.1038/s41598-021-03286-9},
  url = {https://doi.org/10.1038/s41598-021-03286-9}
}

@article{orts2023quantum,
  author = {Orts, Francisco and Filatovas, Ernestas and Garz{\'o}n, Ester M. and Ortega, Gloria},
  title = {A Quantum Circuit to Generate Random Numbers Within a Specific Interval},
  journal = {EPJ Quantum Technology},
  volume = {10},
  number = {1},
  pages = {17},
  year = {2023},
  doi = {10.1140/epjqt/s40507-023-00174-1},
  url = {https://doi.org/10.1140/epjqt/s40507-023-00174-1}
}

@article{sinai2024q,
  author = {Sinai, Nday Kabulo and In, Hoh Peter},
  title = {{Q-RTOP}: Quantum-Secure Random Transaction Ordering Protocol for Mitigating Maximal Extractable Value Attacks in Blockchains With a Priority Gas-Fee Policy},
  journal = {IEEE Access},
  volume = {12},
  pages = {10036--10046},
  year = {2024},
  doi = {10.1109/ACCESS.2024.3351830},
  url = {https://doi.org/10.1109/ACCESS.2024.3351830}
}

@article{elaraby2022quantum,
  author = {Elaraby, Ahmed},
  title = {Quantum Medical Images Processing Foundations and Applications},
  journal = {IET Quantum Communication},
  volume = {3},
  number = {4},
  pages = {201--213},
  year = {2022},
  doi = {10.1049/qtc2.12049},
  url = {https://doi.org/10.1049/qtc2.12049}
}

@article{le2011flexible,
  author = {Le, Phuc Q. and Dong, Fangyan and Hirota, Kaoru},
  title = {A Flexible Representation of Quantum Images for Polynomial Preparation, Image Compression, and Processing Operations},
  journal = {Quantum Information Processing},
  volume = {10},
  number = {1},
  pages = {63--84},
  year = {2011},
  doi = {10.1007/s11128-010-0177-y},
  url = {https://doi.org/10.1007/s11128-010-0177-y}
}

@article{decoodt2023hybrid,
  author = {Decoodt, Pierre and Liang, Tan Jun and Bopardikar, Soham and Santhanam, Hemavathi and Eyembe, Alfaxad and Garcia-Zapirain, Begonya and Sierra-Sosa, Daniel},
  title = {Hybrid Classical--Quantum Transfer Learning for Cardiomegaly Detection in Chest X-Rays},
  journal = {Journal of Imaging},
  volume = {9},
  number = {7},
  pages = {128},
  year = {2023},
  doi = {10.3390/jimaging9070128},
  url = {https://doi.org/10.3390/jimaging9070128}
}

@article{reka2024exploring,
  author = {Reka, S. Sofana and Karthikeyan, H. Leela and Shakil, A. Jack and Venugopal, Prakash and Muniraj, Manigandan},
  title = {Exploring Quantum Machine Learning for Enhanced Skin Lesion Classification: A Comparative Study of Implementation Methods},
  journal = {IEEE Access},
  volume = {12},
  pages = {104568--104584},
  year = {2024},
  doi = {10.1109/ACCESS.2024.3434681},
  url = {https://doi.org/10.1109/ACCESS.2024.3434681}
}

@article{guha2025resq,
  author = {Guha, Dibyasree and Kuiry, Somenath and Mitra, Shyamali and Bhattacharyya, Siddhartha and Das, Nibaran},
  title = {{ResQ}: A Hybrid Classical-Quantum Model for Efficient Breast Cancer Image Classification},
  journal = {Applied Soft Computing},
  volume = {183},
  pages = {113631},
  year = {2025},
  doi = {10.1016/j.asoc.2025.113631},
  url = {https://doi.org/10.1016/j.asoc.2025.113631}
}

@article{sabaawi2024exploiting,
  author = {Sabaawi, Abdulbasit M. A. and Almasaoodi, Mohammed R. and Imre, S{\'a}ndor},
  title = {Exploiting {OFDM} Method for Quantum Communication},
  journal = {Quantum Information Processing},
  volume = {23},
  number = {7},
  pages = {256},
  year = {2024},
  doi = {10.1007/s11128-024-04465-z},
  url = {https://doi.org/10.1007/s11128-024-04465-z}
}

@article{almasaoodi2024novel,
  author = {Almasaoodi, Mohammed R. and Sabaawi, Abdulbasit M. A. and El Gaily, Sara and Imre, S{\'a}ndor},
  title = {A Novel Quantum Orthogonal Frequency-Division Multiplexing Transmission Scheme},
  journal = {International Journal of Advanced Computer Science and Applications},
  volume = {15},
  number = {5},
  pages = {442--449},
  year = {2024},
  doi = {10.14569/IJACSA.2024.0150544},
  url = {https://doi.org/10.14569/IJACSA.2024.0150544}
}

@article{urgelles2024quantum,
  author = {Urgelles, Helen and Garcia-Roger, David and Monserrat, Jose F.},
  title = {Quantum-Based Maximum Likelihood Detection in {MIMO-NOMA} Systems for {6G} Networks},
  journal = {Quantum Reports},
  volume = {6},
  number = {4},
  pages = {533--549},
  year = {2024},
  doi = {10.3390/quantum6040036},
  url = {https://doi.org/10.3390/quantum6040036}
}

@inproceedings{10628250,
  author = {Pathak, Param and Oad, Vidhi and Prajapati, Aditya and Innan, Nouhaila},
  title = {Resource Allocation Optimization in {5G} Networks Using Variational Quantum Regressor},
  booktitle = {2024 International Conference on Quantum Communications, Networking, and Computing (QCNC)},
  pages = {101--105},
  year = {2024},
  publisher = {IEEE},
  doi = {10.1109/QCNC62729.2024.00025},
  url = {https://doi.org/10.1109/QCNC62729.2024.00025}
}

@article{schwabe2025,
  author = {Schwabe, Mierk and Pastori, Lorenzo and de Vega, In{\'e}s and Gentine, Pierre and Iapichino, Luigi and Lahtinen, Valtteri and Leib, Martin and Lorenz, Jeanette Miriam and Eyring, Veronika},
  title = {Opportunities and Challenges of Quantum Computing for Climate Modeling},
  journal = {Environmental Data Science},
  volume = {4},
  pages = {e35},
  year = {2025},
  doi = {10.1017/eds.2025.10010},
  url = {https://doi.org/10.1017/eds.2025.10010}
}

@inproceedings{munasinghe2024qml_climate,
  author = {Munasinghe, Thilanka and Lai, Phung and Wei, Jennifer C. and Hendler, James A. and Cornell, Kimberly A.},
  title = {Assessment of Quantum {ML} Applicability for Climate Actions: Comparison of the Variational Quantum Classifier and the Quantum Support Vector Classifier with Classical {ML} Models},
  booktitle = {2024 IEEE International Conference on Big Data (BigData)},
  pages = {4357--4366},
  year = {2024},
  publisher = {IEEE},
  doi = {10.1109/BigData62323.2024.10825181},
  url = {https://doi.org/10.1109/BigData62323.2024.10825181}
}

@inproceedings{maheshwari2025qcnn,
  author = {Maheshwari, Danyal and Pelzer, Julia and Schulte, Miriam},
  title = {Predicting Heat Plume Temperature and Spatial Location Using Quantum Convolutional Neural Networks},
  booktitle = {2025 International Conference on Quantum Communications, Networking, and Computing (QCNC)},
  pages = {623--627},
  year = {2025},
  publisher = {IEEE},
  doi = {10.1109/QCNC64685.2025.00103},
  url = {https://doi.org/10.1109/QCNC64685.2025.00103}
}

@misc{maheshwari2026qcnn,
  author = {Maheshwari, Danyal and Pelzer, Julia and Schulte, Miriam},
  title = {Quantum Convolutional Neural Networks for Groundwater Heat Plume Prediction: A Surrogate Modeling Approach},
  year = {2026},
  eprint = {2606.23411},
  archivePrefix = {arXiv},
  primaryClass = {quant-ph},
  doi = {10.48550/arXiv.2606.23411},
  url = {https://arxiv.org/abs/2606.23411}
}

@inproceedings{koretsky2021adapting,
  author = {Koretsky, Samantha and Gokhale, Pranav and Baker, Jonathan M. and Viszlai, Joshua and Zheng, Honghao and Gurung, Niroj and Burg, Ryan and Paaso, Esa Aleksi and Khodaei, Amin and Eskandarpour, Rozhin and Chong, Frederic T.},
  title = {Adapting Quantum Approximation Optimization Algorithm ({QAOA}) for Unit Commitment},
  booktitle = {2021 IEEE International Conference on Quantum Computing and Engineering (QCE)},
  pages = {181--187},
  year = {2021},
  publisher = {IEEE},
  doi = {10.1109/QCE52317.2021.00035},
  url = {https://doi.org/10.1109/QCE52317.2021.00035}
}

@inproceedings{mahroo2022hybrid,
  author = {Mahroo, Reza and Kargarian, Amin},
  title = {Hybrid Quantum-Classical Unit Commitment},
  booktitle = {2022 IEEE Texas Power and Energy Conference (TPEC)},
  year = {2022},
  publisher = {IEEE},
  doi = {10.1109/TPEC54980.2022.9750763},
  url = {https://doi.org/10.1109/TPEC54980.2022.9750763}
}

@misc{farhi2014,
  author = {Farhi, Edward and Goldstone, Jeffrey and Gutmann, Sam},
  title = {A Quantum Approximate Optimization Algorithm},
  year = {2014},
  eprint = {1411.4028},
  archivePrefix = {arXiv},
  primaryClass = {quant-ph},
  doi = {10.48550/arXiv.1411.4028},
  url = {https://arxiv.org/abs/1411.4028}
}

@article{zhou2021noise,
  author = {Zhou, Yifan and Zhang, Peng},
  title = {Noise-Resilient Quantum Machine Learning for Stability Assessment of Power Systems},
  journal = {IEEE Transactions on Power Systems},
  volume = {38},
  number = {1},
  pages = {475--487},
  year = {2023},
  doi = {10.1109/TPWRS.2022.3160384},
  url = {https://doi.org/10.1109/TPWRS.2022.3160384}
}

@book{markowitz2000mean,
  author = {Markowitz, Harry M. and Todd, G. Peter},
  title = {Mean-Variance Analysis in Portfolio Choice and Capital Markets},
  publisher = {John Wiley \& Sons},
  year = {2000},
  isbn = {9781883249755},
  url = {https://books.google.com/books?id=eJ8QUsgfZ8wC}
}

@misc{QPO,
  author = {{Qiskit Community}},
  title = {{Qiskit Finance: Portfolio Optimization}},
  year = {2024},
  note = {Qiskit Finance 0.4.1 documentation. Accessed 28 August 2026},
  url = {https://qiskit-community.github.io/qiskit-finance/tutorials/01_portfolio_optimization.html}
}

@article{buonaiuto2023best,
  author = {Buonaiuto, Giuseppe and Gargiulo, Francesco and De Pietro, Giuseppe and Esposito, Massimo and Pota, Marco},
  title = {Best Practices for Portfolio Optimization by Quantum Computing, Experimented on Real Quantum Devices},
  journal = {Scientific Reports},
  volume = {13},
  number = {1},
  pages = {19434},
  year = {2023},
  doi = {10.1038/s41598-023-45392-w},
  url = {https://doi.org/10.1038/s41598-023-45392-w}
}

@article{carrascal2023backtesting,
  author = {Carrascal, Gin{\'e}s and Hernamperez, Paula and Botella, Guillermo and del Barrio, Alberto},
  title = {Backtesting Quantum Computing Algorithms for Portfolio Optimization},
  journal = {IEEE Transactions on Quantum Engineering},
  volume = {5},
  pages = {1--20},
  year = {2024},
  doi = {10.1109/TQE.2023.3337328},
  url = {https://doi.org/10.1109/TQE.2023.3337328}
}

@article{carrascal2024differential,
  author = {Carrascal, Gin{\'e}s and Roman, Beatriz and del Barrio, Alberto and Botella, Guillermo},
  title = {Differential Evolution {VQE} for Crypto-Currency Arbitrage. Quantum Optimization with Many Local Minima},
  journal = {Digital Signal Processing},
  volume = {148},
  pages = {104464},
  year = {2024},
  doi = {10.1016/j.dsp.2024.104464},
  url = {https://doi.org/10.1016/j.dsp.2024.104464}
}

@misc{yalovetzky2021nisq,
  author = {Yalovetzky, Romina and Minssen, Pierre and Herman, Dylan and Pistoia, Marco},
  title = {{NISQ-HHL}: Portfolio Optimization for Near-Term Quantum Hardware},
  year = {2021},
  eprint = {2110.15958},
  archivePrefix = {arXiv},
  primaryClass = {quant-ph},
  doi = {10.48550/arXiv.2110.15958},
  url = {https://arxiv.org/abs/2110.15958}
}

@book{best2000implementing,
  author = {Best, Philip},
  title = {Implementing Value at Risk},
  publisher = {John Wiley \& Sons},
  year = {1998},
  isbn = {9780471972051},
  doi = {10.1002/0470013303},
  url = {https://onlinelibrary.wiley.com/doi/book/10.1002/0470013303}
}

@inproceedings{chow2015risk,
  author = {Chow, Yinlam and Tamar, Aviv and Mannor, Shie and Pavone, Marco},
  title = {Risk-Sensitive and Robust Decision-Making: A {CVaR} Optimization Approach},
  booktitle = {Advances in Neural Information Processing Systems 28},
  pages = {1522--1530},
  year = {2015},
  url = {https://proceedings.neurips.cc/paper/2015/hash/64223ccf70bbb65a3a4aceac37e21016-Abstract.html}
}

@misc{QF,
  author = {{Qiskit Community}},
  title = {{Qiskit Finance: Credit Risk Analysis}},
  year = {2024},
  note = {Qiskit Finance 0.4.1 documentation. Accessed 28 August 2026},
  url = {https://qiskit-community.github.io/qiskit-finance/tutorials/09_credit_risk_analysis.html}
}

@article{wilkens2023quantum,
  author = {Wilkens, Sascha and Moorhouse, Joe},
  title = {Quantum Computing for Financial Risk Measurement},
  journal = {Quantum Information Processing},
  volume = {22},
  pages = {51},
  year = {2023},
  doi = {10.1007/s11128-022-03777-2},
  url = {https://doi.org/10.1007/s11128-022-03777-2}
}

@article{dri2023general,
  author = {Dri, Emanuele and Aita, Antonello and Giusto, Edoardo and Ricossa, Davide and Corbelletto, Davide and Montrucchio, Bartolomeo and Ugoccioni, Roberto},
  title = {A More General Quantum Credit Risk Analysis Framework},
  journal = {Entropy},
  volume = {25},
  number = {4},
  pages = {593},
  year = {2023},
  doi = {10.3390/e25040593},
  url = {https://doi.org/10.3390/e25040593}
}

@article{de2023var,
  author = {de Pedro, Luis and Par{\'i}s Murillo, Ra{\'u}l and L{\'o}pez de Vergara, Jorge E. and L{\'o}pez-Buedo, Sergio and G{\'o}mez-Arribas, Francisco J.},
  title = {{VaR} Estimation with Quantum Computing Noise Correction Using Neural Networks},
  journal = {Mathematics},
  volume = {11},
  number = {20},
  pages = {4355},
  year = {2023},
  doi = {10.3390/math11204355},
  url = {https://doi.org/10.3390/math11204355}
}

@article{grossi2022mixed,
  author = {Grossi, Michele and Ibrahim, Noelle and Radescu, Voica and Loredo, Robert and Voigt, Kirsten and von Altrock, Constantin and Rudnik, Andreas},
  title = {Mixed Quantum-Classical Method for Fraud Detection With Quantum Feature Selection},
  journal = {IEEE Transactions on Quantum Engineering},
  volume = {3},
  pages = {1--12},
  year = {2022},
  doi = {10.1109/TQE.2022.3213474},
  url = {https://doi.org/10.1109/TQE.2022.3213474}
}

\end{document}